\documentclass[onecolumn,showpacs,superscriptaddress, 11pt]{revtex4-2}
\usepackage{amsmath,graphicx,hyperref, amssymb,amsthm,amsfonts,braket,color}
\usepackage{bm, amsmath,amssymb,graphicx, subfigure, braket, float}
\usepackage{xcolor}
\usepackage{graphicx}
\usepackage{xcolor}
\usepackage{tikz}
\usepackage[compact]{titlesec}
\hypersetup{
    colorlinks=true,       
    linkcolor=blue,       
    citecolor=blue,       
    urlcolor=blue         
}
\usepackage{xcolor}
\usepackage{color}
\usepackage{hyperref}
\usepackage[left=0.7in, right=0.7in, top=1in, bottom=1in]{geometry}

\begin{document}
\title{Engineering of Hyperentangled Complex Quantum Networks}

\author{Murad Ahmad}
\address{\small Department of Physics, University of Malakand Chakdara, Dir Lower, KP, Pakistan}
\author{Liaqat Ali}\email{liaqatphysicsuom@gmail.com}
\address{\small Department of Physics, University of Malakand Chakdara, Dir Lower, KP, Pakistan}
\address{Department of Physics, Government Degree College Lund Khwar Mardan, KP, Pakistan}
\author{Muhammad Imran} 
\address{\small National Institute of Lasers and Optronics College
    Pakistan Institute of Engineering and Applied Sciences, Nilore, Islamabad, Pakistan}
\author{Rameez-ul Islam} 
\address{\small National Institute of Lasers and Optronics College
    Pakistan Institute of Engineering and Applied Sciences, Nilore, Islamabad, Pakistan}
\author{Manzoor Ikram} 
\address{\small National Institute of Lasers and Optronics College
    Pakistan Institute of Engineering and Applied Sciences, Nilore, Islamabad, Pakistan}

\author{Rafi Ud Din}
\address{\small Department of Physics, COMSATS University Islamabad, Islamabad, Pakistan}

\author{Ashfaq Ahmad}
\address{\small Department of Chemistry, College of Science,
King Saud University, P.O. Box, 2455, Riyadh-11451, Saudi Arabia}

\author{Iftikhar Ahmad}
\address{\small Department of Physics, University of Malakand Chakdara, Dir Lower, KP, Pakistan}
\address{\small Center for Computational Material Sciences,
University of Malakand Chakdara, Dir Lower, KP, Pakistan}

\begin{abstract}
Hyperentangled states are highly efficient and resource
economical. This is because they enhance the quantum information
encoding capabilities due to the correlated engagement of more
than one degree of freedom of the same quantum entity while
keeping the physical resources at their minimum. Therefore,
initially the photonic hyperentangled states have been explored
extensively but the generation and respective manipulation of the
atomic counterpart states are still limited to only few proposals.
In this work, we propose a new and feasible scheme to engineer the
atomic hyperentangled cluster and ring graph states invoking
cavity QED technique for applicative relevance to quantum biology
and quantum communications utilizing the complex quantum networks.
These states are engineered using both external quantized momenta
states and energy levels of neutral atoms under off-resonant and
resonant Atomic Bragg Diffraction (ABD) technique. The study of
dynamical capacity and potential efficiency have certainly
enhanced the range of usefulness of these states. In order to
assess the operational behavior of such states when subjected to a
realistic noise environment has also been simulated, demonstrating
long enough sustainability of the proposed states. Moreover,
experimental feasibility of the proposed scheme has also been
elucidated under the prevailing cavity-QED research scenario.
\end{abstract}
 \maketitle   
    

\section{Introduction}
The development of fast, efficient and secure computational and
communication techniques based on counterintuitive trait of
quantum mechanics have led to the rapidly emerging field of
quantum informatics
\cite{Feynman2018,Einstein1935,Aspect1982,AspectR21982}. In this
regard, quantum state engineering measurement, large number of
qubits scalability and long coherence times are the stringent
prerequisite for the development of practical quantum technologies
\cite{Nielsen2002,Bouwmeester1997}. Now, in the historical
context, the entanglement has played a key role to address the
foundational problems of quantum theory as well as to implement a
multitude quantum information tasks
\cite{Bell1964,Aspect1982,Zidan2020}. It enables us to implement
the efficient and successful quantum data transmission, processing
and manipulation as well as to asses coherence retainment and
losses in the quantum information encoded in a quantum system
\cite{Kenfack2017,Brune1996,Raimond2001}. The rapid development of
various entanglement-based quantum information protocols have
ushered the subject matter to quite a mature stage by now
\cite{Zidan2018,Zhong2020}. Moreover, for multi-party quantum
communication, the quantum information needs to be distributed
among many parties coherently bonded into a complicated entangled
network morphology. Such quantum networks are naturally found in
biological systems and may also potentially become the building
block of quantum internet. These quantum networks have the
capability to store the quantum information on their nodes which
are interlinked through various quantum channels. In this way
quantum information can be effectively and nonlocally distributed
from one place to another making its dissemination quite practical
throughout the whole network with high fidelity \cite{Kimble2008}.

Now entanglement, a fundamental ingredient of quantum networks has
been generated through various experimental techniques by
utilizing the relevant specific degree of freedoms (DOF)
associated with a physical system. Such a coherent quantum
correlation is generally engineered using a single DOF of many
parties i.e. photon's polarization frequency or the spatial modes,
spin of electron, energy levels of an atom or its external momenta
states. Whereas, entanglement of more than one degree of freedom
is commonly referred to as the hyperentangled state for two or
more parties \cite{Kwait2012}. Till now, the photonic systems have
been extensively studied and are assumed to be the major resource
of hyperentangled states generation. However, atomic systems have
also been recently utilized for the hyperentangled states
engineering and manipulation
\cite{Yang2005,Nawaz2017,Nawaz2018,Nawaz2019}. Such atomic
hyperentangled states employ the quantized atomic internal and
external DOFs. It is well known that the addition of an extra DOF
in an entangled state will enhance the quantum communication
channel capacity logarithmically i.e. $2^{n}$ \cite{Barreiro2008},
and hence boost the quantum information tasks such as
hyperentanglement purification \cite{Wang2016}, hyperentanglement
teleportation \cite{Ali2021,Ali2022-1,Ali2022-2,Ali2024,Wang2015},
Bell state analysis \cite{Wei2007} and hyperentanglement
concentration \cite{Li2015} while keeping the physical resource
level at its bare minimum. The atomic hyperentangled states are
generally engineered through Atomic Bragg Diffraction (ABD)
\cite{Nawaz2017,Nawaz2018,Nawaz2019}. It is important to note that
ABD, an experimentally feasible tool, has already been employed to
handle several quantum foundational issues as well as the quantum
state engineering and processing tasks
\cite{Kunze1996,Durr1998,Islam2015,Ikram2015}.

In the present work, we proposed schematics to engineer a number
of hyperentangled graph states with diverse morphologies such as
$n$-partite linear quantized momenta cluster states, 2D cluster
states and ring graph states through Atomic Bragg Diffraction
technique. The cluster states especially 2D-cluster states serve
as the fundamental resource for the measurement based one-way
quantum computing model \cite{Raussendorf2001,Nielsen2006}. This
model encompasses such states as a computational resource and step
by step processes the information through single-qubit
measurements. In this regard, the experimental possibility of this
model has already been tested in a significant way through the
manipulation of $4$-photon cluster state \cite{Waltheri2005}.
Moreover, these states can be utilized for any complex multi-gate
quantum information protocol due to their strong natural
decoherence resistance, stability and inherently deterministic
nature \cite{Anis2021}. Keeping in view the vitality and
versatility of such states, a plethora of proposals regarding
cluster state engineering in solid-state, atomic and photonic
systems through diverse techniques have been put forward in the
recent decades while hinting out their
respective applicability for one-way computing \cite{Browne2005,Tokunaga2005,Vallone2005,Tokunaga2008,Gao2010,Cho2005,Dong2006,Zhang2007,Lee2008,Lee2009,Gonta2009,Ballester2011,Barrett2005,Duan2005,Chen2006,Tanamoto2006,Xue2006,Zhangi2006,Zheng2006,Blythe2006}%
. On practical backgrounds, the linear optic techniques based
schemes have already been demonstrated experimentally
\cite{Waltheri2005,Kiesel2005,Zhang2006}. These techniques are, in
general, probabilistic and become difficult to implement
experimentally with the increase in the state dimensionality and
the subsequent additive nature of net failure probability .
However, the techniques based on cavity QED are inherently
deterministic, the atomic states under realistic experimental
conditions, can produce such graph states with probability of
success approaching unity
\cite{Raussendorf2001,Nielsen2006,Waltheri2005,Briegel2004,Hein2004,IslamR12008,Briegel2001}.
Furthermore, spontaneous emission, a usual constraint on state
coherence in atom-field cavity QED systems do not pose any
significant threat on the dimensionality of the graph states being
engineered, because here we are playing with the off-resonant ABD
that involves only virtual Rabi cycles with no real absorption or
emission process invoked.

The present work is arranged as follows: section-I is a brief
introduction to the problem. Section-II is an overview of the
mathematical tool i.e. ABD for the tagging of atoms with the
cavities in momentum space leading to the subsequent state
engineering. Section-III narrates in detail the engineering of
various hyperentangled graph states. The last section provides a
summary along with the discussion over the envisioned merits of
the presented schematics. This section also elucidates briefly
over the experimental feasibility of the proposed work in the
context of realistic and contemporary research scenario.

\section{Tagging of atoms in momentum space}
\begin{figure}[h!]
\begin{center}
\includegraphics[height=7.7cm,width=8cm]{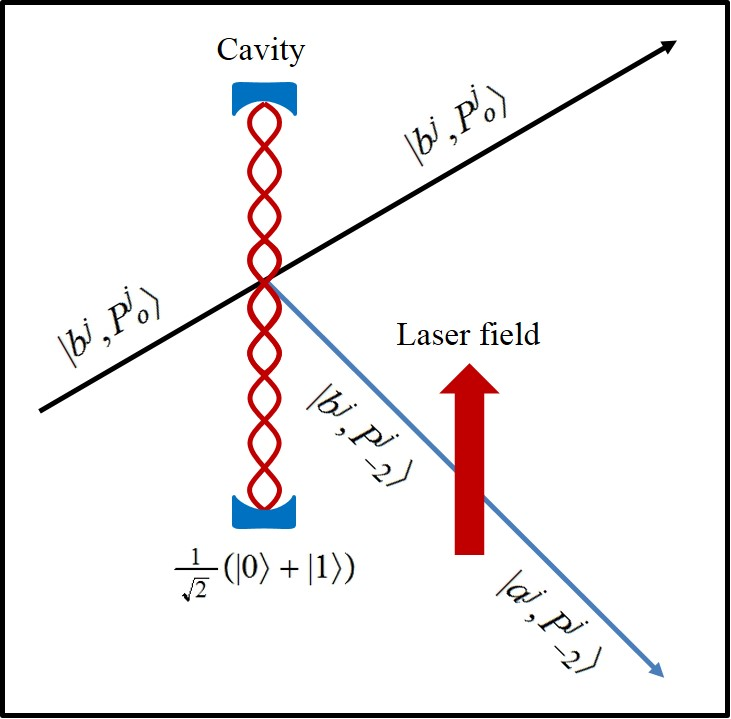}\linespread{0.9}
\end{center}
\vspace*{-7mm}\caption{\fontsize{10}{10}\selectfont{Tagging of type-1 atoms with the cavity.}}\label{Tagging}%
\end{figure}

We consider the two different sets of two-level neutral atoms, one
to engineer the hyperentangled states called type-1 atoms. These
type-1 atoms, apart from the discrete energy levels, are also
taken external quantized transverse momentum. The second type are
auxiliary atoms used to erase the cavities quantum information and
are termed as type-2 atoms. They are considered quantized only in
internal states and are used as the cavity information eraser
through swapping via resonant interaction. The type-2 atoms are
also employed for joining or entangling various independent
quantum states through dispersive atom-field interactions much
similar to the phase gate operation \cite{Zou2007}. We also assume
two high-Q cavities with both being taken initially to be in
{$(|0_{j}\rangle +|1_{j}\rangle )/\sqrt{2}$} photon superposition
states with $j=1,2$. Now the type-1 atoms, initially prepared in
their ground states $\left\vert b_{j}\right\rangle $ with momentum
$\left\vert P_{o_{j}}\right\rangle $, interact off-resonantly with
the cavities under Bragg regime cavity QED scenario as shown in
Fig. \ref{Tagging}. Thus the initial state vector for the first
atom traversing the cavity-1 will be;
\begin{equation}
\left\vert \Psi {(0)}\right\rangle =\frac{1}{\sqrt{2}}{[|0\rangle
+|1\rangle ]}\otimes |b,P_{o}\rangle  \label{A}
\end{equation}
Where $|P_{o}\rangle=\hbar k$ is the quantized transverse atomic
momentum along the cavity axis. The Hamiltonian for such an
interaction under dipole and rotating wave approximation is
\cite{Khosa2004};
\begin{equation}
H=\frac{\hat{P}_{x}^{2}}{2m}+\frac{\hbar \Delta }{2}\sigma
_{z}+\hbar \mu Cos(kx)[\sigma _{ab}\hat{c}+\hat{c}^{\dagger
}\sigma _{ba}]  \label{B}
\end{equation}
Here ${\hat{P}_{x}^{2}/2m}$ marks the kinetic energy along the
axis of quantized atomic external momentum, $\sigma_{ab}=|a\rangle
\langle b|$, $\sigma_{ba}=|b\rangle \langle a|$ and
$\sigma_{z}=|a\rangle \langle a|-|b\rangle \langle b|$ corresponds
to the raising, lowering and inversion atomic operators. Whereas
$|b\rangle$($|a\rangle$) denotes the atomic ground (excited)
state, $\hat{c} (\hat{c^{\dagger}})$ stands for the field
annihilation (creation) operator and $\Delta$ specifies the
atom-field detuning while $\mu$ is the Rabi frequency for such
atom-field interaction. Now the proposed state vector for an
arbitrary interaction time $t$ will be
\cite{Khosa2004,IslamR12008};
\begin{eqnarray}
|\Psi {(t)}\rangle &=&\exp ^{-i(\frac{P_{x}^{2}}{2m\hbar}-\frac{\Delta }{2}%
)t}\sum_{l=-\infty }^{\infty }\left[ C_{(0,b)}^{(P_{l})}\rangle
(t)|0,b,P_{l}\rangle\right.  \nonumber \\
&&\left.+C_{1,b}^{p_{l}}(t)|1,b,P_{l}\rangle
+C_{0,a}^{p_{l}}(t)|0,a,P_{l}\rangle \right]  \label{C}
\end{eqnarray}
The summation over $l$ signifies the accumulative nature of atomic
quantized transverse momentum acquired during the atom-field
interaction \cite{Khosa2004,Khosa2005,Khosa2006,Khan1999}.
Furthermore, $C_{j,k}^{P_{l}}$
represent probability amplitudes for $k = a, b$ and $j = 0, 1$ i.e. for $%
j^{th}$ state of the cavity field and $k^{th}$ atomic state. The
larger value of atom-field detuning considered here is to ensure
virtual Rabi cycles leaving negligible chance for spontaneously
emitted photons, resulting in the guaranteed persistence of
coherence throughout the interaction. Now solving the
Schrodinger's wave equation i.e. $i\hbar\frac{\partial}{\partial
t}\left\vert \Psi_{AB}(t)\right\rangle=H\left\vert
\Psi_{AB}(t)\right\rangle$, using adiabatic approximation i.e,
$\Delta \gg \omega _{r}\gg \mu ^{2}/{\Delta }$ yields the state
vector as;
\begin{equation}
|\Psi {(t)}\rangle =\frac{1}{\sqrt{2}}[|0,b,P_{o}\rangle
+|1,b,P_{-2}\rangle ]  \label{D}
\end{equation}
Here $|P_{o}\rangle$ and $|P_{-2}\rangle$ are the two mutually
orthogonal split atomic momenta wavepackets yielded by
off-resonant first order ABD when the interaction time is taken
as, $t=2\pi\Delta/\mu^2$
\cite{Khosa2004,Khosa2005,Khosa2006,Khan1999,Khalique2003,Islam2007,Islam2008,Qamar2003,Islam2009}.
For further mathematical details on quantum state engineering
through ABD, one is referred to these well cited resources
\cite{Islam2008,Islam2009,Ikram2015,Nawaz2017,Nawaz2018,Nawaz2019}.
The above equation shows that the atom is now entangled in only
one degree of freedom i.e. atomic external quantized momentum
$|P_o\rangle (|P_{-2}\rangle)$ with the cavity-field. However, two
spatially well separated split atomic wavepackets with ground
states atomic signatures are travelling in mutually perpendicular
directions rendered permissible under first order Bragg's
diffraction after leaving the cavity. Thus we can expose any one
momenta component to a laser beam for a pre-calculated time
selected while keeping in view the classical longitudinal momentum
or velocity of the atom. Under this semiclassical
interaction scenario, the atom will be excited from the ground state $|b\rangle$ to the excited state $%
|a\rangle$ after interacting with the laser field. Such an
atom-field interaction is governed by the semiclassical
interaction picture Hamiltonian as \cite{Scully1997};
\begin{equation}
H^{sc} =\frac{\hbar\Omega_{R}}{2}[\sigma _{ab}+\sigma _{ba}]
\label{E}
\end{equation}
Where, $\Omega _{R}=\left\vert \wp _{ba}\right\vert \varepsilon
/\hbar $ is Rabi frequency with $\varepsilon$ describing the
amplitude of applied classical field and $\wp _{ba}$ represents
the transition dipole matrix element. Moreover, $\sigma
_{ab}(\sigma _{ba})$ stands for atomic raising (lowering) operator
operating under the due action of the classically applied field.
Now solving the Schrodinger's wave equation and adjusting the
interaction time to $t_{sc}=\pi/\Omega _{R}$, the final state
vector comes to be;
\begin{equation}
|\Psi {(t_{sc})}\rangle =\frac{1}{\sqrt{2}}[|0,b,P_{o}\rangle
-i|1,a,P_{-2}\rangle ]  \label{F}
\end{equation}
It is evident that Ramsey laser field was applied only to
$|P_{-2}\rangle$ component of the split atomic momenta wavepacket
along Y-axis, the axis across which atomic momentum is again
treated. The above equation stands for a hyperentangled state of
two-level type-1 atom coherently correlated with the cavity field.
The same procedure can be extended to engineer the $n$-partite
hyperentangled state. For this purpose we pass a stream of
two-level identical type-1 atoms through the cavity one by one
under off-resonant ABD. This will consequently generate a state of
the form expressed in eq. (\ref{D}) but now with an entangled and
extended Hilbert space. Then one of their spatially well separated
momenta components interact with the classical laser fields under
similar conditions as employed for the first atom. Therefore, the
$n$-partite hyperentangled state becomes;
\begin{equation}
|\Psi\rangle
=\frac{1}{2^{n/2}}\prod_{j=1}^{n}[|0,b^{j},P_{o}^{j}\rangle+(i)^{3j}|1,a^{j},P_{-2}^{j}\rangle
]  \label{G}
\end{equation}
Here the superscript $j$ varies from $1$ to $n$ and stands for the
number of identical atoms interacting with the cavity field
consecutively, one after the other. This number $n$ can be quite
large in the range of thousands. This is because the experimental
feasibility of such a large number of consecutive interactions has
already been demonstrated within the permissible time limits of
the onset of any appreciable decoherence
\cite{Haroche2006,Haroche2020}.
\section{Hyperentangled graph states engineering}
\subsection{Generation of hyperentangled cluster states}
Expression (\ref{G}) above, for any arbitrarily large $n$,
represents a multipartite GHZ type hyperentangled atomic state
tagged or coherently correlated with the cavity field because all
these atoms have interacted successively with same cavity under
off-resonant ABD. Now, it is evident that in order to engineer a
pure hyperentangled atomic state, we have to disentangle the
cavity field from this expression. More explicitly stated, we have
to erase the cavity information appropriately to yield the
intended multipartite hyperentangled atomic state. Thus in order
to erase the cavity quantum information, we pass the auxiliary
atom i.e. type-2 atom through the cavity resonantly for the
removal of such quantum information. Since these auxiliary atoms
are used to swap and subsequently erase the cavity-field
information interactively for which one needs to invoke only the
energy levels and therefore auxiliary atoms are taken to be moving
with classical momentum. Moreover, the resonant atom should be
considered initially in its ground state $|g\rangle$ while its
excited state is denoted by $|e\rangle$. Thus the initial state
vector for such resonant interaction can be expressed as;
\begin{equation}
|\Psi {(0)}\rangle =\frac{1}{2^{n/2}}\prod_{j=1}^{n}[|0,b^{j},P_{o}^{j}%
\rangle +(i)^{3j}|1,a^{j},P_{-2}^{j}\rangle ]\otimes|g\rangle
\label{H}
\end{equation}
Here the tensor product incorporates the two set of atoms i.e. in
the first product $n$ atoms of type-1 that are appropriately
entangled in momentum and internal degrees with the cavity field,
whereas the outer product of type-2 atom that is needed to
accomplish resonant interaction with the cavity field. As already
stated, this interaction will swap the cavity information onto the
initially ground state i.e. $|g\rangle$ atom which will be
subsequently erased via Ramsey field interactions. The interaction
causing the cavity quantum information swapping is governed by the
resonant atom-field interaction Hamiltonian given as under
\cite{Scully1997};
\begin{equation}
H=\hbar \mu [\sigma _{eg}\hat{c}+\hat{c}^{\dagger }\sigma _{ge}]
\label{I}
\end{equation}
Here $\sigma _{eg}=|e\rangle \langle g|(\sigma _{ge}=|g\rangle
\langle e|)$ denotes the atomic raising (lowering) operators and
$\mu$ is the atom-field coupling constant. This resonant
atom-field interaction will result into the disentanglement of the
cavity field from the entangled atoms. At this stage, the
auxiliary atom will be energy-level wise entangled with the type-2
quantized momenta atoms. Hence auxiliary atom swaps the cavity
information into its internal levels and leaves the cavity into
vacuum state i.e. $|0\rangle$ which can be traced out of the
expression. This auxiliary atom is then employed for joining
various building block states to engineer the desired graph state
through resonant and dispersive interactions. Hence, in this way
we can generate various type of cluster states and most general
graph states of any arbitrary dimension with morphology
encapsulating many nodes and edges. Prior to furnishing few easy
but important examples in details to illustrate the case, it is
worth clarifying here that the state expressed in eq. (\ref{H})
should be taken as generic in the context of dimensionality and we
can truncate it to express a hyper-superposition, with cavity
information duly erased, of just a single atom to any higher
dimensional hyperentangled atomic state, when and where required.

\subsection{Bipartite hyperentangled linear cluster states}
\begin{figure}[h!]
\begin{center}
\includegraphics[height=10.5cm,width=15.5cm]{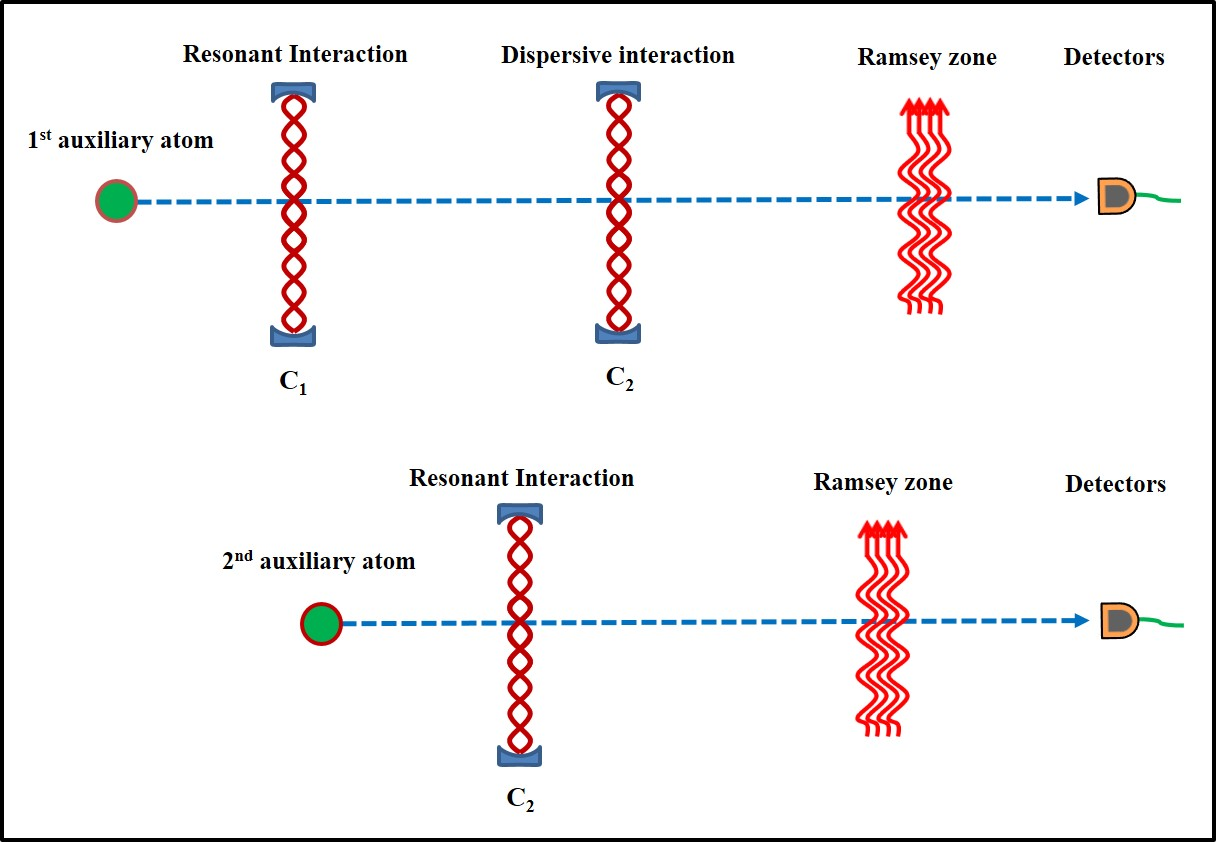}\linespread{0.5}
\end{center}\vspace*{-7mm}\caption{\fontsize{10}{10}\selectfont{The diagram represents the interaction of auxiliary
atoms with the cavity-1 and cavity-2. Here $\left\vert
g\right\rangle $ and Here $\left\vert e\right\rangle $
explore the ground and excited states of auxiliary atoms. Whereas, $C_1$ and $C_2$ stand for the cavity-1 and 2 respectively.}}\label{cavities}%
\end{figure}

For bipartite system we consider two type-1 atoms which are
already tagged with their respective cavities i.e. cavity-1 and
cavity-2 and two type-2 auxiliary atoms. Such an initial state may
be expressed as follows;
\begin{equation}
|\Phi^{1}{(0)}\rangle
=\frac{1}{\sqrt{2}}[|0_{1},b^{1},P_{o}^{1}\rangle-i|1_{1},a^{1},P_{-2}^{1}\rangle
]\otimes|g_{1}\rangle\nonumber
\end{equation} and
\begin{equation}
|\Phi^{2}{(0)}\rangle
=\frac{1}{\sqrt{2}}[|0_{2},b^{2},P_{o}^{2}\rangle-i|1_{2},a^{2},P_{-2}^{2}\rangle
]\otimes|g_{2}\rangle \label{I-2}
\end{equation}
The tensor product state therefore comes to be;
\begin{eqnarray}
|\Psi {(t_{1}=0)}\rangle  &=&\left\vert \Phi^{1}(0)\right\rangle
\otimes
\left\vert \Phi^{2}(0)\right\rangle   \nonumber \\
&=&\frac{1}{2}\left[ {|0}_{{1}},{0}_{{2}},{b^{1},b^{2},P_{o}^{1},P_{o}^{2}%
\rangle -i|{0}_{{1}},1_{{2}},{b^{1},a^{2},P_{o}^{1},P_{-2}^{2}}\rangle }%
\right.   \nonumber \\
&&\left.
-i{|1_{1},{0}_{{2}},{a^{1},}b^{2},{P_{-2}^{1},}P_{o}^{2}\rangle
-|1_{1},1_{2},{a^{1},}a_{2},{P_{-2}^{1},}P_{-2}^{2}\rangle
}\right] \otimes {|g}_{{1}},{g}_{{2}}\rangle \label{I-3}
\end{eqnarray}
 Now we pass $1^{st}$ type-2 atom in its ground state $|g_{1}\rangle$,
 resonantly through cavity-1 and then dispersively through the
 cavity-2. Quantum information of the cavity-1 is thus swapped to
 this auxiliary atom and it leaves the cavity in vacuum state $|0_{1}\rangle$.
 This resonant interaction is governed by the Hamiltonian [eq. (\ref{I})] mentioned earlier.
 Then the same auxiliary atom, after swapping cavity-1 information via resonant interaction, enters into the cavity-2.
 Here in cavity-2, the auxiliary atom engages itself in dispersive interaction described by the Hamiltonian
 expressed in the coming paragraph. This dispersive interaction impart a field-dependent phase to the state
 in accordance with specifically selected interaction time. After that, we pass the second type-2 atom which is again
initially prepared in its ground state $|g_{2}\rangle$ from the
cavity-2 resonantly that swaps the cavity information and leaves
the cavity-2 in vacuum field state while transferring the cavity
state information to the atom as was the case of first auxiliary
atom. The resonant interaction time for both auxiliary atoms will
be $\pi $-Rabi cycle i.e. ${t_{1}=\pi /2\mu }$. Finally, to obtain
the hyperentangled bipartite linear cluster state of type-1 atoms,
we pass both the auxiliary atoms through the Ramsey field prior to
the state-selective atomic detection immediately after leaving
these cavities. Ramsey interactions are a stringent pre-requisite
here because they erase the information carried by the auxiliary
atoms. This is because the Ramsey's interaction leads to the
transformations \cite{Scully1997};
\begin{equation}
\left\vert g_{j}\right\rangle \longrightarrow
\frac{1}{\sqrt{2}}\left[ \left\vert g_{j}\right\rangle +\left\vert
e_{j}\right\rangle \right] \nonumber
\end{equation}
\begin{equation}
\left\vert e_{j}\right\rangle \longrightarrow
\frac{1}{\sqrt{2}}\left[ \left\vert g_{j}\right\rangle -\left\vert
e_{j}\right\rangle \right] \label{J}
\end{equation}
The sequential interaction of these auxiliary atoms with the
fields of the cavity is shown in the Fig. \ref{cavities}. Here we
give a detailed step-wise description. The $1^{st}$ auxiliary atom
resonantly interacts with the cavity-1. Such a resonant atom-field
interaction will swap the cavity-1's information. Thus the cavity
is left in vacuum field state i.e. $|0_{1}\rangle$ that can be
simply traced out from the final equation. This atom-field
resonant interaction is governed by the Hamiltonian i.e. eq.
(\ref{I}). Hence, setting an interaction time $t_{1}=\pi /2\mu $,
the initial state vector i.e. eq. (\ref{I-3}) becomes;
\begin{eqnarray}
|\Psi {(t_{1})}\rangle &=&\frac{1}{2}\left[
{|b^{1},g^{1},P_{o}^{1}\rangle
-|a^{1},e^{1},P_{-2}^{1}\rangle }\right] \nonumber \\
&&{\otimes } \left[ {|0_{2},b^{2},P_{o}^{2}\rangle -i|1_{2},a_{2},P_{-2}^{2}\rangle }%
\right] \otimes {|g^{2}}\rangle  \label{K}
\end{eqnarray}%
Now in the next step, the auxiliary atom-1, emerging from cavity-1
then dispersively interacts with cavity-2 and generates the
desired entanglement by imparting local phase depending over
amplitude of the cavity. This dispersive interaction is governed
by the Hamiltonian \cite{Zubairy2004};
\begin{equation}
H_{d}=\hbar \lambda \left[ \hat{c}\hat{c}^{\dagger }|e\rangle \langle e|-%
\hat{c}^{\dagger }\hat{c}|g\rangle \langle g|\right]
\label{Des.H}
\end{equation}%
Where $\lambda =\mu _{d}^{2}/\Delta $ is the effective Rabi frequency, $%
\Delta $ is the detuning and $\mu _{d}$ denotes the coupling
parameter. This type of dispersive interaction has already been
experimentally demonstrated in the context of quantum phase gate
\cite{Rauschenbeutel1999}. In our case it will assist to furnish
the residual part of the Ising interaction joining the two nodes.
Therefore, the final state vector after dispersive interaction of
auxiliary atom-1 with the cavity-2 for any arbitrary interaction time $t_{2}$ becomes;%
\begin{eqnarray}
|\Psi \left( t_{2}\right) \rangle  &=&\frac{1}{2}\left[ {%
|0_{2},b^{1},b^{2},g^{1},P_{o}^{1},P_{o}^{2}\rangle }\right.   \nonumber \\
&&\left. {-i}\exp \left( -i\lambda t_{2}\right) {%
|1_{2},b^{1},a^{2},g^{1},P_{o}^{1},P_{-2}^{2}\rangle }\right.   \nonumber \\
&&\left. +\exp \left( -2i\lambda t_{2}\right) {%
|0_{2},a^{1},b^{2},e^{1},P_{-2}^{1},P_{o}^{2}\rangle }\right.   \nonumber \\
&&\left. {{-i}\exp \left( i\lambda t_{2}\right)
|1_{2},a^{1},a^{2},e^{1},P_{-2}^{1},P_{-2}^{2}\rangle }\right]
\otimes |g^{2}\rangle   \label{L}
\end{eqnarray}
Now the auxiliary atom-2 which is initially prepared in its ground
state $ \left\vert g^{2}\right\rangle $ resonantly interacts with
cavity-2 for an interaction time {$t_{3}=\pi /$}$2\mu _{r}$ in
order to swap the cavity information as already mentioned above.
Such an interaction is governed by the resonant interaction
picture Hamiltonian given in eq. (\ref{I}). Therefore, using the
initial conditions suggested by eq. (\ref{L}), the final state
vector after this resonant interaction lasting for time
{$t_{3}=\pi /$}$2\mu _{r}$ yields the following state vector;
\begin{eqnarray}
|\Psi (t_{3})\rangle  &=&\frac{1}{2}\left[ {%
|b^{1},b^{2},g^{1},g^{2},P_{o}^{1},P_{o}^{2}\rangle }\right.   \nonumber \\
&&\left. {-\exp \left( -i\lambda t_{2}\right) \
|b^{1},a^{2},g^{1},e^{2},P_{o}^{1},P_{-2}^{2}\rangle }\right.   \nonumber \\
&&\left. {+\exp \left( -2i\lambda t_{2}\right) {\
|a^{1},b^{2},e^{1},g^{2},P_{-2}^{1},P_{o}^{2}\rangle }}\right.   \nonumber \\
&&\left. {{-\exp \left( i\lambda t_{2}\right) }%
|a^{1},a^{2},e^{1},e^{2},P_{-2}^{1},P_{-2}^{2}\rangle }\right]
\label{M}
\end{eqnarray}
Thus in this way the cavity field information is effectively
swapped through the auxiliary atom-2 and leaves the cavity-2 in
vacuum field state $\left\vert 0_{2}\right\rangle $, which is
traced out from the above equation. Now, to engineer the desired
hyperentangled state, we pass both these auxiliary atoms through
the Ramsey zone after emerging out from the respective cavities
prior to detection. This completes the eraser mechanism. Ramsey
zone acts a symmetric Hadamard gate and will coherently split the
atomic internal state in accordance with the transformations
mentioned previously. Further, for the sake of symmetric,
equally-weighted state engineering, the dispersive interaction
time $t_{2}$ carried previously as a running variable is now being
fixed to $t_{2}=\pi/\lambda$. Therefore, utilizing Ramsey
transformations in eq. (\ref{M}), the equal weighted atomic
hyperentangled linear cluster states engineered in accordance with
the state selective detection pattern of the auxiliary atoms come
to be;
\begin{eqnarray}
|\Phi \rangle &=&\frac{1}{2}\left[ \frac{1}{2}\left\{ \left\vert
b^{1},b^{2},P_{o}^{1},P_{o}^{2}\right\rangle +{\left\vert
b^{1},a^{2},P_{o}^{1},P_{-2}^{2}\right\rangle }\right. \right.  \nonumber \\
&&\left. \left. {+\left\vert
a^{1},b^{2},P_{-2}^{1},P_{o}^{2}\right\rangle
+\left\vert a^{1},a^{2},P_{-2}^{1},P_{-2}^{2}\right\rangle }\right\} {%
\otimes \left\vert g^{1},g^{2}\right\rangle }\right.  \nonumber \\
&&\left. \frac{1}{2}\left\{ \left\vert
b^{1},b^{2},P_{o}^{1},P_{o}^{2}\right\rangle -{\left\vert
b^{1},a^{2},P_{o}^{1},P_{-2}^{2}\right\rangle }\right. \right.  \nonumber \\
&&\left. \left. {+\left\vert
a^{1},b^{2},P_{-2}^{1},P_{o}^{2}\right\rangle
-\left\vert a^{1},a^{2},P_{-2}^{1},P_{-2}^{2}\right\rangle }\right\} {%
\otimes \left\vert g^{1},e^{2}\right\rangle }\right.  \nonumber \\
&&\left. \frac{1}{2}\left\{ \left\vert
b^{1},b^{2},P_{o}^{1},P_{o}^{2}\right\rangle +{\left\vert
b^{1},a^{2},P_{o}^{1},P_{-2}^{2}\right\rangle }\right. \right.  \nonumber \\
&&\left. \left. -{\left\vert
a^{1},b^{2},P_{-2}^{1},P_{o}^{2}\right\rangle
-\left\vert a^{1},a^{2},P_{-2}^{1},P_{-2}^{2}\right\rangle }\right\} {%
\otimes \left\vert e^{1},g^{2}\right\rangle }\right.  \nonumber \\
&&\left. \frac{1}{2}\left\{ \left\vert
b^{1},b^{2},P_{o}^{1},P_{o}^{2}\right\rangle -{\left\vert
b^{1},a^{2},P_{o}^{1},P_{-2}^{2}\right\rangle }\right. \right.  \nonumber \\
&&\left. \left. -{\left\vert
a^{1},b^{2},P_{-2}^{1},P_{o}^{2}\right\rangle
+\left\vert a^{1},a^{2},P_{-2}^{1},P_{-2}^{2}\right\rangle }\right\} {%
\otimes \left\vert e^{1},e^{2}\right\rangle }\right]  \label{O}
\end{eqnarray}%
It is evident from the above expression that for any atomic state
detection permutation, we get a hyperentangled atomic linear
cluster state. Further, form the procedure cited above, one can
equally engineer any arbitrarily large $n$-partite atomic
hyperentangled cluster state by coherently joining the states,
taken initially from eq. (\ref{H}) for large enough $n$, through
the above explained Ising interaction.
\subsection{Four-partite hyperentangled 2D-Cluster State}
\begin{figure}[h!]
\begin{center}
\includegraphics[height=7.5cm,width=16cm]{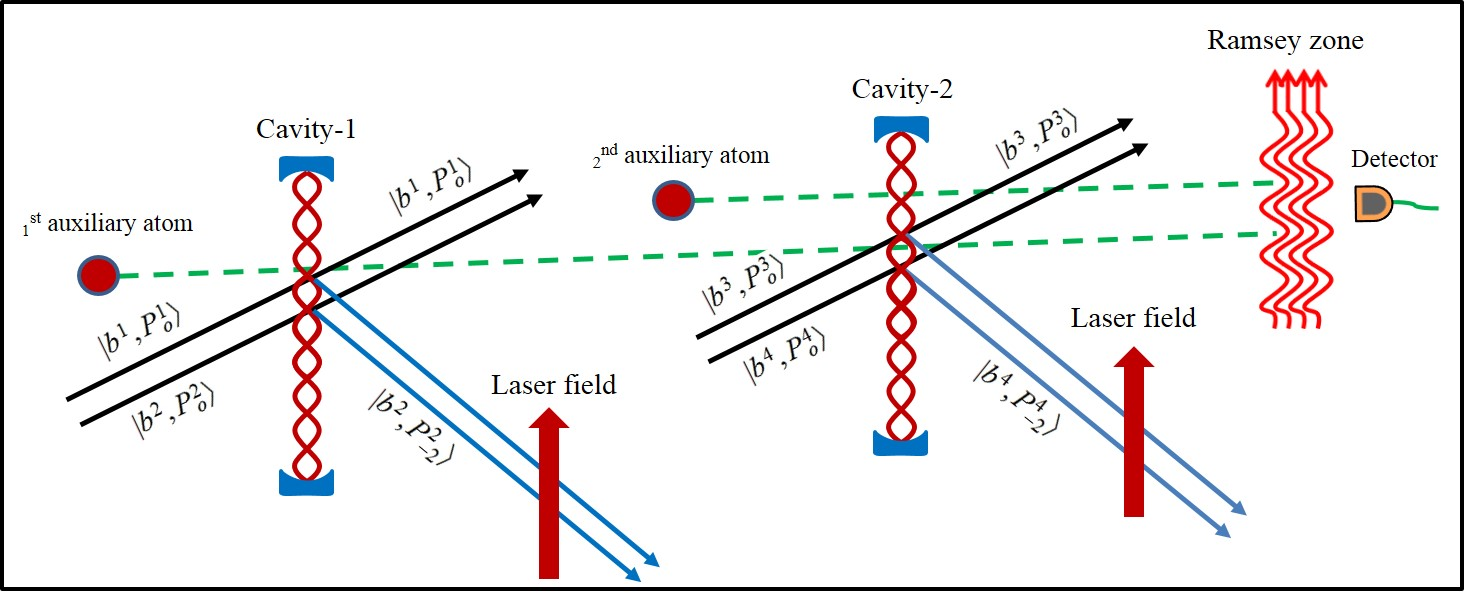}\linespread{0.5}
\end{center}\vspace*{-7mm}\caption{\fontsize{10}{10}\selectfont{The sketch elucidates the
complete process of hyperentangled four-partite 2D-cluster state
engineering. Here $1^{st}$ and $2^{nd}$ atom are off-resonantly
Bragg diffracted from the cavity-1 and the interaction culminates
into generating an entangled state of the form expressed in eq.
(\ref{P}) with the number fixed to $n=2$ in eq. (\ref{H}).
Whereas, the $3^{rd}$ and $4^{th}$ atom traverse off-resonantly
through the cavity-2 in the same manner. Then the first auxiliary
atom interacts resonantly with the cavity-1 and then dispersively
with the cavity-2. Moreover, the $2^{nd}$ auxiliary atom passes
through the cavity-2 resonantly. Finally, these auxiliary atoms
pass through
the Ramsey zone for atomic state-selective detection.}}\label{2D-cluster}%
\end{figure}

Following the previously explained procedure, here we give the
schematics to generate the four-partite hyperentangled atomic
2D-cluster state. It is the most vital case because the
counterpart photonic state was utilized for the experimental
demonstration of one-way computing model \cite{Waltheri2005}. Now,
in order to engineer a four-partite hyperentangled atomic
2D-cluster state we consider four two-level type-1 atoms in which
two of them say $1^{st}$ and $2^{nd}$ atom will pass through the
cavity-1 and the remaining two atoms, say $3^{rd}$ and $4^{th}$
atom, will pass through the cavity-2 under off-resonant Bragg
regime. These two-level neutral atoms are used to generate
cavity-field entanglement as described earlier. In the next step,
we assume two auxiliary atoms say type-2 atoms, which interact
resonantly through their respective cavities to erase the cavities
information in such a way that atom-1 after swapping information
from cavity-1 through resonant interaction passes through cavity-2
where it interacts dispersively. Such dispersive interaction
effectively form a controlled phase gate and thus entangle the two
mutually independent states i.e. initially correlated state of
atom-1, atom-2 and atom-3, atom-4. Therefore, the interaction ends
up with the generation of the desired atomic hyperentangled state
duly completed when these atoms traverse the Ramsey zone and are
subsequently detected via state-selective detectors. These
schematics are as illustrated in Fig. \ref{2D-cluster}.\\

 For the desired building block engineering as
explained earlier, the $1^{st}$ and $2^{nd}$ type-1 atoms pass
through the cavity-1 off-resonantly under atomic Bragg diffraction
domain one after the other for an interaction time $t=2\pi \Delta
/\mu ^{2}$. This coherently splits the atomic momenta wavepackets
into two equal parts i.e. with equal probability amplitudes. After
that one component of each atom is exposed to the
classical laser field under the same conditions as proposed for eq. (\ref{F}%
) to generate hyperentanglement. This procedure produces a pair of
the GHZ type entangled states as follows;
\begin{equation}
|\Psi ^{1,2}(t)\rangle =\frac{1}{\sqrt{2}}\left[
|0_{1},b^{1},b^{2},P_{o}^{1},P_{o}^{2}\rangle
+|1_{1},a^{1},a^{2},P_{-2}^{1},P_{-2}^{2}\rangle \right]
\label{P}
\end{equation}%
Similarly, the $3^{rd}$ and $4^{th}$ atom will pass through the
cavity-2 off-resonantly repeating all the procedural steps as
mentioned for $1^{st}$ and $2^{nd}$ atoms ends up with the
generation of a similar doubly tagged or entangled state as
follows;
\begin{equation}
|\Psi ^{3,4}(t)\rangle =\frac{1}{\sqrt{2}}\left[
|0_{2},b^{3},b^{4},P_{o}^{3},P_{o}^{4}\rangle
+|1_{2},a^{3},a^{4},P_{-2}^{3},P_{-2}^{4}\rangle \right]
\label{Q}
\end{equation}%
The next step is to erase the cavities information through the
auxiliary atoms which will leave the cavities in their respective
vacuum states and hence will disentangle the cavities which are
left into respective vacuum state and hence can be traced out.
Such auxiliary atoms are initially taken in their ground states
$\left\vert g^{1}\right\rangle $ and $\left\vert
g^{2}\right\rangle $ respectively, as depicted in the Fig.
\ref{2D-cluster}. Therefore, the initial product state, from eq.
(\ref{P}) and eq. (\ref{Q}) may be expressed as follows;
\begin{equation}
|\Psi^{2D} (0) \rangle =|\Psi ^{1,2}(t)\rangle {\otimes }|\Psi
^{3,4}(t)\rangle {\otimes \left\vert g^{1},g^{2}\right\rangle }
\label{R}
\end{equation}%
Now the first auxiliary atom will interact resonantly for a
complete $\pi$-Rabi cycle through the cavity-1 and then interacts
dispersively with the the cavity-2 for an interaction time
$t_{d}$. After that, the second auxiliary atom will pass through
the cavity-2 resonantly, again for an interaction time equal to
$\pi$-Rabi cycle. In this way, both the cavities's information
gets completely swapped by these atoms. Thus both the cavities are
left in their vacuum field states after such atom-field
interactions. Therefore, the initial state vector i.e. eq.
(\ref{R}) after such interactions transforms to;
\begin{eqnarray}
|\Psi ^{2D}(t_{d})\rangle  &=&\frac{1}{2}\left[
|g^{1},g^{2},b^{1},b^{2},b^{3},b^{4},P_{o}^{1},P_{o}^{2},P_{o}^{3},P_{o}^{4}%
\rangle \right.   \nonumber \\
&&\left. +i\exp \left( -i\lambda t_{d}\right)
|g^{1},e^{2},b^{1},b^{2},a^{3},a^{4},P_{o}^{1},P_{o}^{2},P_{-2}^{3},P_{-2}^{4}\rangle
\right.
\nonumber \\
&&\left. +\exp \left( -2i\lambda t_{d}\right)
|e^{1},g^{2},a^{1},a^{2},b^{3},b^{4},P_{-2}^{1},P_{-2}^{2},P_{o}^{3},P_{o}^{4}\rangle
\right.
\nonumber \\
&&\left. -i{\exp \left( i\lambda t_{d}\right) }%
|e^{1},e^{2},a^{1},a^{2},a^{3},a^{4},P_{-2}^{1},P_{-2}^{2},P_{-2}^{3},P_{-2}^{4}\rangle %
\right]   \label{S}
\end{eqnarray}
With the cavity fields $|0_{1}$ and $|0_{2}$ duly traced out of
the above expression. Now, as stated in the previous section,
these auxiliary atoms need to be disentangled from the above
expression through Bell-basis measurement. For this purpose these
auxiliary atoms are passed through the Ramsey zone immediately
after emerging from the cavities. Such Ramsey zone enacts Hadamard
transforms to the atomic internal states of the auxiliary atoms
according eq. (\ref{J}) prior to the state detection. Therefore,
the final state vector carrying the state-selective detection
pattern of the auxiliary atoms i.e. $\left\vert
g^{1},g^{2}\right\rangle$, $\left\vert g^{1},e^{2}\right\rangle$,
$\left\vert e^{1},g^{2}\right\rangle$ and $\left\vert
e^{1},e^{2}\right\rangle$, may be expressed as;
\begin{eqnarray}
|\Psi ^{2D}\rangle  &=&\frac{1}{2}\left[ \frac{1}{2}\left\{ \left\vert {%
b^{1},b^{2},b^{3},b^{4},P_{o}^{1},P_{o}^{2},P_{o}^{3},P_{o}^{4}}%
\right\rangle \right. \right.   \nonumber \\
&&\left. \left. +i\exp \left( -i\lambda t_{d}\right) \left\vert {%
b^{1},b^{2},a^{3},a^{4},P_{o}^{1},P_{o}^{2},P_{-2}^{3},P_{-2}^{4}}%
\right\rangle \right. \right.   \nonumber \\
&&\left. \left. +\exp \left( -2i\lambda t_{d}\right) \left\vert a{%
^{1},a^{2},b^{3},b^{4},P_{-2}^{1},P_{-2}^{2},P_{o}^{3},P_{o}^{4}}%
\right\rangle \right. \right.   \nonumber \\
&&\left. \left. -i{\exp \left( i\lambda t_{d}\right) }\left\vert a{%
^{1},a^{2},a^{3},a^{4},P_{-2}^{1},P_{-2}^{2},P_{-2}^{3},P_{-2}^{4}}%
\right\rangle \right\} {\otimes \left\vert
g^{1},g^{2}\right\rangle }\right.
\nonumber \\
&&\left. +\frac{1}{2}\left\{ \left\vert {%
b^{1},b^{2},b^{3},b^{4},P_{o}^{1},P_{o}^{2},P_{o}^{3},P_{o}^{4}}%
\right\rangle \right. \right.   \nonumber \\
&&\left. \left. -i\exp \left( -i\lambda t_{d}\right) \left\vert {%
b^{1},b^{2},a^{3},a^{4},P_{o}^{1},P_{o}^{2},P_{-2}^{3},P_{-2}^{4}}%
\right\rangle \right. \right.   \nonumber \\
&&\left. \left. +\exp \left( -2i\lambda t_{d}\right) \left\vert a{%
^{1},a^{2},b^{3},b^{4},P_{-2}^{1},P_{-2}^{2},P_{o}^{3},P_{o}^{4}}%
\right\rangle \right. \right.   \nonumber \\
&&\left. \left. +i{\exp \left( i\lambda t_{d}\right) }\left\vert a{%
^{1},a^{2},a^{3},a^{4},P_{-2}^{1},P_{-2}^{2},P_{-2}^{3},P_{-2}^{4}}%
\right\rangle \right\} {\otimes \left\vert
g^{1},e^{2}\right\rangle }\right.
\nonumber \\
&&\left. +\frac{1}{2}\left\{ \left\vert {%
b^{1},b^{2},b^{3},b^{4},P_{o}^{1},P_{o}^{2},P_{o}^{3},P_{o}^{4}}%
\right\rangle \right. \right.   \nonumber \\
&&\left. \left. +i\exp \left( -i\lambda t_{d}\right) \left\vert {%
b^{1},b^{2},a^{3},a^{4},P_{o}^{1},P_{o}^{2},P_{-2}^{3},P_{-2}^{4}}%
\right\rangle \right. \right.   \nonumber \\
&&\left. \left. -\exp \left( -2i\lambda t_{d}\right) \left\vert a{%
^{1},a^{2},b^{3},b^{4},P_{-2}^{1},P_{-2}^{2},P_{o}^{3},P_{o}^{4}}%
\right\rangle \right. \right.   \nonumber \\
&&\left. \left. +i{\exp \left( i\lambda t_{d}\right) }\left\vert a{%
^{1},a^{2},a^{3},a^{4},P_{-2}^{1},P_{-2}^{2},P_{-2}^{3},P_{-2}^{4}}%
\right\rangle \right\} {\otimes \left\vert
e^{1},g^{2}\right\rangle }\right.
\nonumber \\
&&\left. +\frac{1}{2}\left\{ \left\vert {%
b^{1},b^{2},b^{3},b^{4},P_{o}^{1},P_{o}^{2},P_{o}^{3},P_{o}^{4}}%
\right\rangle \right. \right.   \nonumber \\
&&\left. \left. -i\exp \left( -i\lambda t_{d}\right) \left\vert {%
b^{1},b^{2},a^{3},a^{4},P_{o}^{1},P_{o}^{2},P_{-2}^{3},P_{-2}^{4}}%
\right\rangle \right. \right.   \nonumber \\
&&\left. \left. -\exp \left( -2i\lambda t_{d}\right) \left\vert a{%
^{1},a^{2},b^{3},b^{4},P_{-2}^{1},P_{-2}^{2},P_{o}^{3},P_{o}^{4}}%
\right\rangle \right. \right.   \nonumber \\
&&\left. \left. -i{\exp \left( i\lambda t_{d}\right) }\left\vert a{%
^{1},a^{2},a^{3},a^{4},P_{-2}^{1},P_{-2}^{2},P_{-2}^{3},P_{-2}^{4}}%
\right\rangle \right\} {\otimes \left\vert
e^{1},e^{2}\right\rangle }\right] \label{T}
\end{eqnarray}
Hence, the above equation represents our desired four partite
hyperentangled 2D-cluster states for example, if the detectors
record ${\left\vert e^{1},g^{2}\right\rangle }$, then the
engineered equal-weighted four-partite hyperentangled 2D-cluster
state can be expressed as;
\begin{eqnarray}
|\Psi ^{^{e^{1},g^{2}}}\rangle  &=&\frac{1}{2}\left[ \left\vert {%
b^{1},b^{2},b^{3},b^{4},P_{o}^{1},P_{o}^{2},P_{o}^{3},P_{o}^{4}}%
\right\rangle \right.   \nonumber \\
&&\left. +\left\vert {%
b^{1},b^{2},a^{3},a^{4},P_{o}^{1},P_{o}^{2},P_{-2}^{3},P_{-2}^{4}}%
\right\rangle \right.   \nonumber \\
&&\left. +\left\vert a{%
^{1},a^{2},b^{3},b^{4},P_{-2}^{1},P_{-2}^{2},P_{o}^{3},P_{o}^{4}}%
\right\rangle \right.   \nonumber \\
&&\left. -\left\vert a{%
^{1},a^{2},a^{3},a^{4},P_{-2}^{1},P_{-2}^{2},P_{-2}^{3},P_{-2}^{4}}%
\right\rangle \right]   \label{XX}
\end{eqnarray}
In expression (\ref{XX}), we have taken $t_{d}=\pi/2\lambda$
instead of running variable for the sake of symmetry and
simplicity. This is standard hyperentangled 2D-cluster atomic
state whereas the rest of the equally probable states can be
transformed into the desired standard form through local unitary
operations or gates.
\subsection{Hyperentangled Ring Graph State Engineering}
Graph states represent the most generalized form for the
multipartite entangled states including the commonly known states
like Bell states, W-states and GHZ states. These specific cases
can be engineered unitarily manipulating the relevant graph states
through local operations. Graph states may incorporate a large
number of qubits that can be binded into entangled correlations
with diverse morphologies and geometries having multiple node
connections among various graph edges. The ring graph state is a
consecutively connected close ring of entangled qubits and
generation of such states is vitally important for the
construction of a quantum network that can be used for fast,
secure and distributive quantum communication \cite{Cohen2018}. In
this section, we propose a scheme \ to engineer a hyperentangled
atomic ring graph state. It is based on $n$-number of type-1
two-level neutral atoms which interact off-resonantly with their
respective cavities under Bragg regime and the interaction
culminates into the generation of the entangled states of the type
given in eq. (\ref{G}). In the next step, an auxiliary two-level
neutral atom i.e. type-2 atom, in its ground state $\left\vert
g\right\rangle $ will
interact dispersively through the cavity-1 for an interaction time $\pi $%
-Rabi cycle. The same auxiliary atom is then passed through all of
the other cavities, where it again interacts dispersively forming
effectively a controlled phase gate through each atom-cavity field
interaction and such interactions thus result in the generation of
entanglements among the various cavities and their respective
Bragg diffracted atoms. Now to connect the first and last cavity
i.e. to produce the ring, we pass this same atom again through
cavity-1 dispersively. After that such auxiliary atom is passed
through the Ramsey zone for detection. This completes the
procedure for the engineering of cavity field-Bragg diffracted
atoms ring graph state. Finally we erase the quantum information
carried by each cavity by passing $n$-auxiliary atoms, where one
atom interacts resonantly with one specific cavity and swaps the
field contained in the cavity. All these type-2 atoms, after
passing through their specific cavities traverse Ramsey's zones
and, are detected through state-selective detectors. This results
into the generation of $n$-partite atomic ring graph state which
can easily be transformed into a hyperentangled graph by exposing
various sites with a classical laser beam for a pre-calculated
time.

\begin{figure}[ht]
\begin{center}
\includegraphics[height=10cm,width=11.5cm]{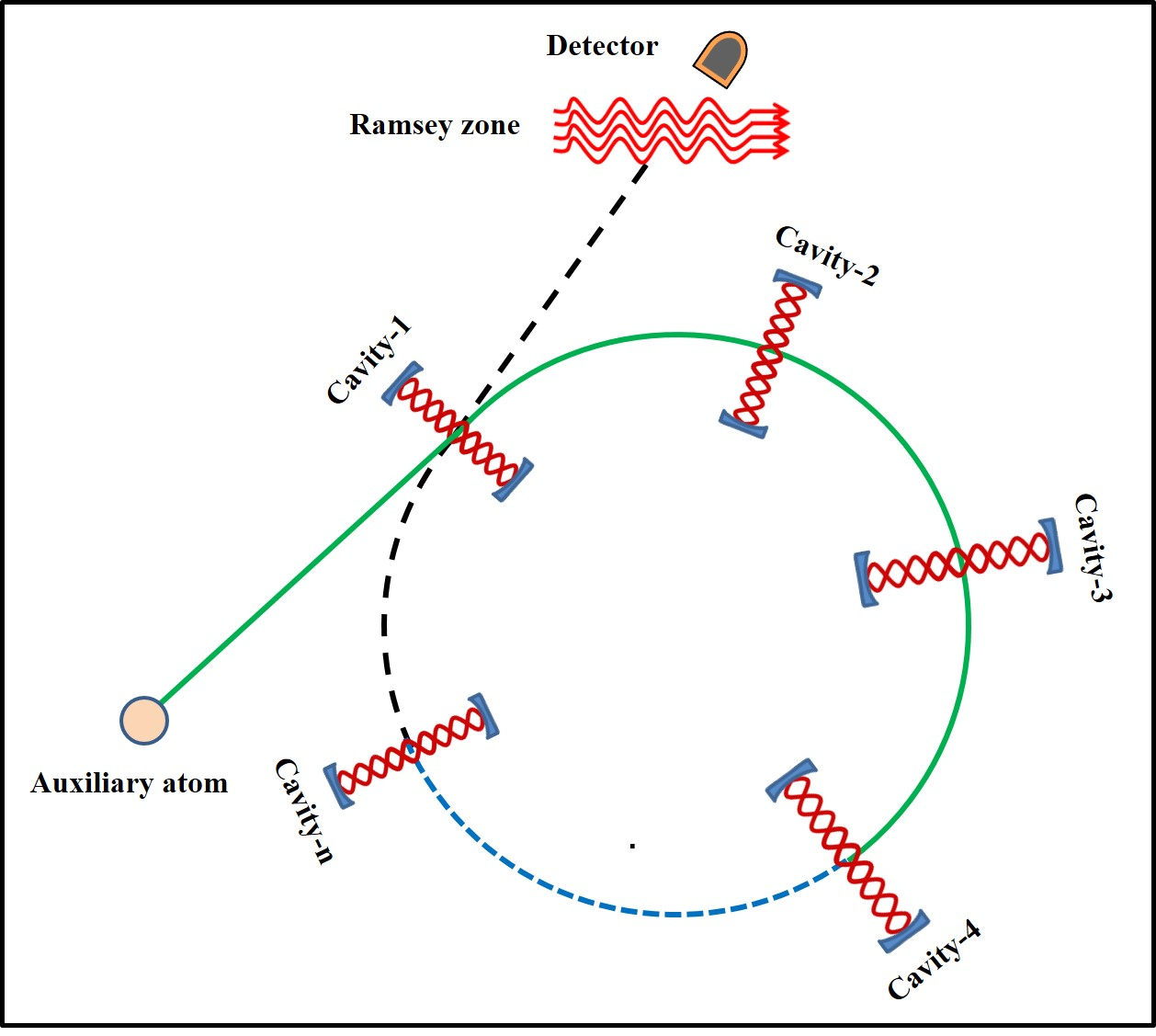}\linespread{0.5}
\end{center}\vspace*{-7mm}\caption{\fontsize{10}{10}\selectfont{The schematics explores the complete
mechanism of atomic hyperentangled ring graph state engineering.}}\label{Ring}%
\end{figure}

Below we present the mathematical details for the
engineering of such a state in a self explanatory manner. We start
from the entangled state i.e. eq. (\ref{D}) engineered between the
cavity-field and the off-resonantly Bragg diffracted atom from the
cavity field. In the next step we pass an auxiliary atom which is
initially prepared in its ground state $\left\vert g\right\rangle
$ through the cavity-1 dispersively. Thus the initial state vector
for such an atom-field interaction may be engineered as follows;
\begin{equation}
\left\vert \Phi ^{^{1}}(t_{1}=0)\right\rangle
=\frac{1}{\sqrt{2}}[\left\vert 0_{1},b^{1},P_{\circ
}^{1}\right\rangle -i\left\vert
1_{1},a^{1},P_{-2}^{1}\right\rangle ]\otimes \left\vert
g\right\rangle \label{U}
\end{equation}%
The Hamiltonian governing such a dispersive interaction is expressed in eq. (\ref%
{Des.H}). Thus, solution of the Schrodinger wave equation for an
arbitrary interaction time $t$ yields the state vector as;
\begin{equation}
\left\vert \Phi ^{^{1}}(t_{1})\right\rangle
=\frac{1}{\sqrt{2}}\left[ \left\vert 0_{1},g,b^{1},P_{\circ
}^{1}\right\rangle -i{\exp \left( -i\lambda t_{1}\right)
}\left\vert 1_{1},g,a^{1},P_{-2}^{1}\right\rangle \right]
\label{V}
\end{equation}%
After successful completion of this process, the same auxiliary
atom will pass through the second cavity, again in dispersive
regime, to generate the entanglement between these cavities via
controlled phase gate operation while taking $t_{1}=\pi/\lambda$.
Moreover, this atom-field interaction takes place under similar
conditions as described in above equation without involving any
external atomic momenta components because the auxiliary atoms are
assumed to move with classical momentum. Thus the initial state
vector for the second
dispersive interaction i.e. $\left\vert \Phi^{2}(t_{2}=0)\right\rangle =\frac{1}{2}\left[ \left( {|0}%
_{{1}},{b^{1},P_{o}^{1}\rangle +i|1_{1},{a^{1},P_{-2}^{1}}\rangle
}\right)
\otimes \left( {|0}_{{2}},{b^{2},P_{o}^{2}\rangle -i|1_{2},{a^{2},P_{-2}^{2}}%
\rangle }\right) \right] \otimes {|g}\rangle $ transforms into the
following state after dispersive interaction of the auxiliary atom
with the second cavity:
\begin{eqnarray}
\left\vert \Phi ^{^{2}}(t_{2})\right\rangle  &=&\frac{1}{2}\left[
\left\vert 0_{1},0_{2},g,b^{1},b^{2},P_{\circ }^{1},P_{\circ
}^{2}\right\rangle \right.
\nonumber \\
&&\left. +i{\exp \left( -2i\lambda t_{2}\right) }\left\vert
0_{1}1_{2},g,b^{1},a^{2},P_{\circ }^{1}P_{-2}^{2}\right\rangle
\right.
\nonumber \\
&&\left. -i{\exp \left( -i\lambda t_{2}\right) }\left\vert
1_{1},0_{2},g,a^{1},b^{2},P_{-2}^{1},P_{\circ }^{2}\right\rangle
\right.
\nonumber \\
&&\left. +{\exp \left( i\lambda t_{2}\right) }\left\vert
1_{1},1_{2},g,a^{1},a^{2},P_{-2}^{1},P_{-2}^{2}\right\rangle
\right] \label{W}
\end{eqnarray}
Similarly, passing the same auxiliary atom through $n$-number of
cavities dispersively, one after the other, for an interaction time $t=\pi/\lambda$ in each case, the final state vector can be expressed as;%
\begin{equation}
\left\vert \Phi ^{^{n}}(t)\right\rangle =\left( \frac{1}{2}\right) ^{\frac{n%
}{2}}\prod_{j=1}^{n}\left[ |0_{j},b^{j},P_{o}^{j}\rangle
-\prod_{l=1}^{2n-1}(i)^{(2l+1)}|1_{j},a^{j},P_{-2}^{j}\rangle
\right] \otimes \left\vert g\right\rangle   \label{X}
\end{equation}
Now, in order to close the ring, we finally pass the auxiliary
atom through cavity-1 again and invoke the same dispersive
interaction.
Therefore, the above generalized hyperentangled cluster state becomes;%
\begin{equation}
\left\vert \Phi ^{^{(n+1)}}(t)\right\rangle =\left( \frac{1}{2}\right) ^{%
\frac{n+1}{2}}\prod_{j=1}^{n+1}\left[
|0_{j},b^{j},P_{o}^{j}\rangle
-\prod_{l=1}^{2n-1}(i)^{(2l+1)}|1_{j},a^{j},P_{-2}^{j}\rangle
\right] \otimes \left\vert g\right\rangle \label{Z}
\end{equation}%
Furthermore, to swap the cavities information we pass $n$-extra
auxiliary atoms resonantly through all of these cavities
independently such that one resonant atom will interact only with
one cavity. Such resonant atoms should be initially prepared in
their ground states $\left\vert g^{(i)}\right\rangle $, where
$i=1,2,3,...n$. These resonant individual and independent
atom-field interactions are governed by the interaction picture
Hamiltonian given in eq. (\ref{I}). Now these resonant atoms will
swap the information of all cavities and left them in vacuum field
states, respectively which can be traced out from the main
expression. Finally, the resonant atoms carrying the quantum
information will pass through the Ramsey zone and will be detected
in their ground or excited states, respectively. Thus Bell-basis
procedure is invoked to erase the cavity quantum information
swapped and carried by these auxiliary atoms. Such state-selective
detection of the $n$-auxiliary atoms finally yield us with $2^{n}$
equally probable states corresponding to the detection sequence
$\prod\limits_{j=1}^{n}\prod\limits_{k=2}^{n}\left( \alpha
^{j},\beta ^{k}\right) $ with $\left( \alpha ,\beta \right)
=\left( e,g\right) $. Now if these all atoms are, for the sake of
simplified presentation, assumed to be detected in their ground
states then the corresponding engineered atomic
hyperentangled ring graph state may be expressed as;%
\begin{equation}
\left\vert \Phi ^{^{(n+1)}}(t)\right\rangle =\left( \frac{1}{2}\right) ^{%
\frac{n+1}{2}}\prod_{j=1}^{n+1}\left[ |b^{j},P_{o}^{j}\rangle
+|a^{j},P_{-2}^{j}\rangle \right] \label{AB}
\end{equation}%
Note that to engineer a hyperentangled ring graph state, the value
of $j$ must be greater than or equal to $2$ i.e. $j\geqslant 2$.
If $j=1$ then the state should only be a hyper-superposition state
of a single atom. Thus for $j=2$, the corresponding graph state
should follow a triangular geometry. Similarly, for $j=3$, it
should be in the form of a square/rectangle. Therefore, any
desired geometry hyperentangled graph state can be engineered
through appropriate value of $j$.
\subsection{Dynamics of the Engineered states under Realistic Noise Environment}
We have briefly described in the previous section as well as in
the next section, the potentially stable nature of the engineered
states qualitatively based on two very important features of such
states.
\begin{enumerate}
   \item Neutral ground state atoms interact off-resonantly with
    the field implying that there is almost nil chance for state
    decay and decoherence that results from spontaneous emission.
    \item External quantized momenta states are known to be
    decoherence resistant states and are generally treated as the
    most fittest state under the domain of Quantum Darwinism
    \cite{Ball2008}.

\end{enumerate}

After successful engineering of the different types of graph
states, we are now in a position to characterize these states in
term of the decay when they are coupled to a realistic classical
environment. This mechanism will help us to assess the usefulness
of these states for numerous quantum information tasks. It is
evident that the actual potential of various quantum systems
should be gauged in accordance with their respective preservation
of the quantum coherence, entanglement and quantum information for
longer time \cite{RAH1, RAH2}. Thus, a quantum system can be
completely described when it is exposed to an environment.
Therefore, we interact the currently engineered graph states with
the classical fields for the purpose to evaluate the dynamical
capacity of these graph states in the context of preservation of
the entanglement for long enough times. Now keeping the importance
of cluster states in view as well as for the sake of simplicity,
we expose the cluster state i.e. eq. (\ref{O}) to the Stochastic
field expressed by the Hamiltonian \cite{RAH3};
\begin{align}
\begin{split}
H(t)=&H_1(t) \otimes I_2 \otimes I_3 \otimes I_4+I_1 \otimes
H_2(t)\otimes I_3\otimes I_4+\\ &I_1 \otimes I_2 \otimes
H_3(t)\otimes I_4+I_1 \otimes I_2\otimes I_3 \otimes H_4(t).
\end{split}
\label{hmm}
\end{align}
Here, $ H_n(t)=\xi I+\lambda \Delta_n(t)\sigma^x$ explores the
Hamiltonian for a single qubit with $n \in \{1, 2, 3, 4\}$ and $I$
($\sigma^x$) is the identity matrix. Whereas, the $\Delta_n(t)$
designates the Stochastic parameter which flips between $\pm1$.
Moreover, $\lambda$ marks the coupling constant for the given
quantum system and the environment. Now at any time t, the time
evolved state for this system can be computed as
$\rho(t)=U(t)\rho(0)U^{\dagger}(t)$. Where $U(t)=\exp{[-i\int^t_0
H(t)dt}]$ is the unitary time operator. Thus, in order to know
about the nature of the system i.e. whether the system is
separable or entangled, we find out the negative expectation
values of the time evolved matrix as $EW(t)=\hbox{Tr}[R_O
\rho(t)]$, where $R_O=\frac{1}{2}I-\vert W\rangle \langle W \vert$
is the witness operator \cite{RAH3,RAH4}. According to this
criterion, the state will be separable for $EW(t) \leq 0$ and will
be entangled for $EW(t)>0$.

\begin{figure}[h!]
\begin{center}
\includegraphics[height=9cm,width=11.5cm]{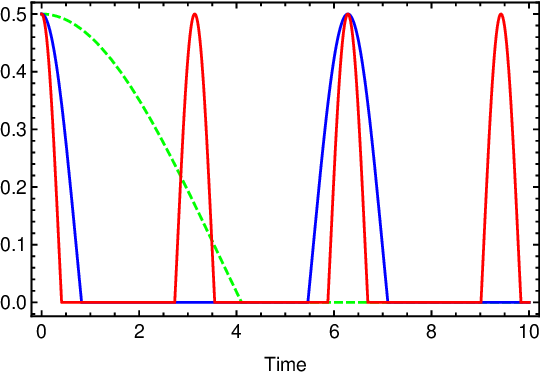}\linespread{0.5}
\end{center}\vspace*{-7mm}\caption{\fontsize{10}{10}\selectfont{The dynamical map
of the engineered graph state coupled with the classical random
field when $\lambda=0.1$ (green dashed line), $\lambda=0.5$ (Blue
line), $\lambda=1$ (red line) with
$\Delta_1=\Delta_2=\Delta_3=\Delta_4=1$ against the unitless time
parameter $t=10$.}}\label{EWO}%
\end{figure}

Fig. \ref{EWO} explores the dynamical capacity of the currently
engineered graph state when coupled with the random field. It is
clear from the graph that the state remains entangled
 for the whole interaction time. Thus, for $\lambda=0$ i.e. minimum interaction
 level, the state is maximally entangled but with the increase of
 the interaction between the system and environment, the state
 losses its entanglement very rapidly. Moreover, the sudden
 entanglement death and birth revivals are observed for any value of $\lambda$
 suggesting the transformation of the engineered graph state into
 a free or disentangled state as well as re-mergence of the entanglement. Thus, due to potential efficiency, the
 system quickly converts itself back into an entangled state
 \cite{Kenfack2017,Ali2023}. This inter-conversion of the graph state between the resource regime
 and free state occurs periodically as long as the system
 exposed to the classical field which concludes that the system
 effectively stays entangled for an infinite time. The plot also shows that the
 amplitude of entanglement does not vary with the time suggesting
 that the system regains its originality of the encoded
 entanglement. Therefore, from the above discussion, we can safely conclude
 that all of the currently engineered graph states are good
 resources for various quantum information protocols such as
 quantum networks and quantum computations. This claim, apart from the
 evaluated cluster state, is justified because all the state engineering is carried out via similar controlled phase gate operations

\section{Experimental feasibility and Conclusion}
The subject of Quantum information is now entering into more
complex era comprised of multipartite quantum entangled webs with
different graph morphologies \cite{Elliott2002,Simon2017}. Recent
explorations have amply demonstrated that quantum coherence
prevailing among complex, multiqubit entangled system that
generally fall into one type or the other of graph states can
potentially explain many phenomena, and hitherto taken to be
intractable
\cite{Lloyd2011,Lambert2013,Gadiyaram2019,Kim2021,Cohen2018}.
Therefore, such high dimensional, multipartite states and their
indepth dynamical exploration is vitally important for the
realization of quantum communication networks, distributed quantum
computing and the operational understanding of the biological
phenomena
\cite{McCutcheon2016,Cavalcanti2015,Liao2018,Dynes2019,Ball2011,Arndt2009,Cohen2018,Cirac1999}.
Cavity QED is a fundamental technique used to handle many quantum
information tasks under ABD criteria
\cite{Haroche2006,Haroche2020,Cortinas2020,Meystre2021}. However,
the experimental feasibility for the realization of such
multipartite quantum networks is posed with a crucial problem i.e.
the risk of decoherence. The main problem in generating these
states can be subdivided into two major domains.
\begin{enumerate}
    \item The engineering of hyperentangled atomic building block
    states i.e. state of the type expressed in eq. (\ref{F}), mainly through off-resonant ABD.
    \item The establishment of coherent connections among numerous graph edges and nodes
    with the help of Ising interactions carried mainly through dispersive
    atom-field interactions.
\end{enumerate}
Now, for a long enough graph state a lot of atom-field
interactions have to be sought out. So, as generally expected, the
rate of decoherence should increase with the number of
interactions seemingly making the present scenario apparently
intractable experimentally. However, the situation is not that
bleak in reality. This is because we use neutral atoms having
quantized external momenta states that interact off-resonantly
with the cavity field in our schematics. The advantage of these
quantized momenta states of neutral atoms is that they minimize
the threat of decoherence i.e almost nil chance of spontaneously
emitted photons during these interactions. In fact these atomic
quantized momenta states are recognized for their high resistance
against decoherence and such states are considered to be the
fittest states under what is known as Quantum Darwinism
\cite{Ball2008}. Moreover, the experimental feasibility of the
interaction of large number of atoms i.e. in the range of
thousands has already been demonstrated experimentally through
high-Q cavities. Thus Quantum Darwinism guarantees the successful
implementation and utilization of such states upto arbitrarily
large number of atom-field interactions for longer enough times
\cite{Deleglise2008,Gleyzes2007}. Furthermore, the experimental
demonstration of Bragg diffraction of atoms through classical as
well as quantized fields have also been performed experimentally
upto $8^{th}$ order resulting in a significantly good spatial
separations, efficiency and fringe visibility
\cite{Kunze1996,Gleyzes2007,Giltner1995,Martin1087,Vernooy1998}.
Another state of the art experimental demonstration of the Bragg
diffraction of $^{85}Rb$ atoms through a laser field of wavelength
$780$ $nm$ has also been reported with good overall results
\cite{Durr1998}. Their employed working parameters nicely follows
the criterion of the atomic Bragg diffraction and are summarized
as follows; $M=85$ $amu$ is the mass of the $^{85}Rb$ atom, $\mu
=2\pi \times 16.4$ $MHz$ is the vacuum Rabi frequency, $\lambda
=780$ $nm$ is the wavelength of the applied laser field and
$\omega _{r}=\hbar k^{2}/2M=2.4\times 10^{4}$ $rad/s$ is the
recoil frequency of the atoms. Moreover, the finesse of the cavity
is $F=4.4\times 10^{5}$ and the atom-field detuning is $\Delta
=1GHz$. As stated, these parameters are in good agreement with the
off-resonant ABD criteria i.e $\omega _{r}+\Delta \gg \mu
\sqrt{n_{c}}/2$. The most important aspect of in this context is
the atom-field interaction time which is $0.5\mu s$ and this is
much smaller than the life-time of the cavity that is as high as
fraction of a second
\cite{Durr1996,Munstermann1999,Munstermanni1999,Puppe2004}. On the
other hand, one can also use helium atoms to experimentally
demonstrate
the atomic Bragg diffraction related state engineering with $M=4$ $amu$, $\omega _{r}=1.06$ $MHz$, $%
\lambda =543.5$ $nm$, $\Delta =6.28$ $GHz$, $\mu ^{2}/4\Delta =120$ $kHz$, $%
F=7.85\times 10^{6}$ and $t=13$ $\mu s$ \cite{Khosa2004,Hood2001}.
Therefore, keeping the aforementioned experimental work in view,
we are quite optimistic about the experimental execution of our
work. Moreover, the proposed schematics are expected to yield good
fidelities. It is important to note that the fidelities of cavity
QED based atom-field interactions are hampered mainly by the
interaction time errors encountered during such interactions.
However, due to being long interaction time regime with spatially
well separated outputs, ABD is generally not affected by such
minor errors and using cold atom samples taken from
Magneto-Optical Traps (MOT) further reduces such a stringent
source of errors because cold atoms have almost no velocity spread
\cite{Kunze1996,Giltner1995,Martin1087,Vernooy1998}.

In summary, we have presented an experimentally executable scheme
to engineer different type of hyperentangled cluster states along
with the most general graph states which can be potentially
employed for quantum communicational networks, distributed quantum
computation as well as for understanding and simulating many
complex phenomena including molecular dynamics, coherence effects
in long, intricate quantum multipartite chains and biological
complexity prevailing in both zoological and botanical domains
e.g. birds navigation and photosynthesis. It is worth noting here
that Ising interaction comprised basically of dispersive and
Ramsey interactions is generic in nature and can be easily
employed to construct any envisionable graph morphology with
arbitrarily large dimensionality. As already stated, such quantum
states will play significant role in the future framework of
quantum information along with the deeper knowledge of the working
dynamics of various biological phenomena \cite{Mohseni2014}.
Moreover, as suggested by the invoked strategy and schematics, one
can engineer any multipartite quantum network having diverse
structural morphologies when and wherever needed for the study of
any complicated, high dimensional quantum phenomenon. Similarly
the exploration of these complex multipartite structures will
enhance the operational knowledge of the holistic nature of
quantum theory through multipartite coherence and interference
effects \cite{Aharonov2018}. In this work we have also shown that
how one can hyper-superpose $n$-number of atoms connected with
their respective cavities using off-resonant ABD. Here the
cavities are considered as nodes i.e. the focal point from where
we can inject, transfer and distribute the quantum information
over an extended coherence pertaining zone. In addition, we have
engineered a generalized $n$-nodes hyperentangled graph state by
connecting cavities in the form of a ring. Such engineered state
has its own figure of merits for quantum networks. Finally, we
have discussed the experimental feasibility of our proposed scheme
in light of the prevailing realistic laboratory parameters which
guarantee the experimental demonstration of our proposed scheme.
\section*{Declarations}
\section*{Acknowledgement}
I. A. acknowledges Center for Computational Materials Science,
University of Malakand, Pakistan for providing resources for
conducting this research project. A. A. acknowledges Researchers
Supporting Project, King Saud University, Riyadh, Saudi Arabia
under grant number RSPD2024R666.
\subsection*{Ethical statement}
 All the results are original and analyzed analytically. The paper is neither submitted nor published in any other journal.
\subsection*{Declaration of competing interest}
The authors declare that they have no known competing financial
interests or personal relationships that could have appeared to
influence the work reported in this paper.
\subsection*{Data availability statement}
The data generated and/or analyzed during the current study are
not publicly available for legal/ethical reasons but are available
from the corresponding author on reasonable request.

\protect


\begin{thebibliography}{9}                                                                                                %
\section*{REFERENCES}
\bibitem{Einstein1935} A. Einstein, B. Podolsky, N. Rosen, ``Can Quantum-Mechanical Description of Physical Reality Be Considered Complete?" Phys. Rev.
\textbf{47}, 777-780 (1935).

\bibitem{Feynman2018} R. P. Feynman, ``Simulating physics with computers" Int. j. Theor. phys. \textbf{21}, 6/7
(2018).

\bibitem{Aspect1982} A. Aspect, P. Grangier, G. Roger, ``Experimental realization of Einstein-Podolsky-Rosen-Bohm Gedanken experiment: A new violation of Bell's
inequalities" Phys. Rev. Lett. \textbf{49}(2), 91-94 (1982).

\bibitem{AspectR21982} A. Aspect, J. Dalibard, G. Roger,
``Experimental Test of Bell's Inequalities Using Time-Varying
Analyzers" Phys. Rev. Lett. \textbf{49}(25), 1804-1807 (1982).

\bibitem{Nielsen2002} M. A. Nielsen, I. L. Chuang, ``Quantum Computation and Quantum Information" Cambridge University
Press, (London) (2002).

\bibitem{Bouwmeester1997} D. Bouwemeester, J. W. Pan, K. Mattle, M. Eibl, H. Weinfurter, A. Zeilinger "Experimental quantum teleportation" Nature,
\textbf{390}(660), 575-579 (1997).


\bibitem{Bell1964} J. S. Bell, ``On the Einstein podolsky rosen paradox" Physics \textbf{1}, 195-200 (1964).


\bibitem{Zidan2020} M. Zidan, ``A novel quantum computing model based on entanglement degree" Mod. Phys. Lett. B \textbf{34}, 2050401 (2020).

\bibitem{Kenfack2017} L. T. Kenfack, M. Tchoko, G. C. Fouokeng, L. C.
Fai, ``Dynamics of tripartite quantum correlations in mixed
classical environments: the joint effects of the random telegraph
and static noises" Int. J. Quantum Inf. \textbf{15}, 1750038
(2017).

\bibitem{Raimond2001} J. M. Raimond, M. Brune, S. Haroche, ``Manipulating quantum entanglement with atoms and photons in a cavity" Rev. Mod. Phys. \textbf{73}
565 (2001).

\bibitem{Brune1996} M. Brune, E. Hagley, J. Dreyer, X. Maitre,
A. Maali, C. Wunderlich, J. M. Raimond, and S. Haroche,
"Observing the Progressive Decoherence of the Meter in a
Quantum Measurement" Phys. Rev. Lett. \textbf{77}(24), 4887-4890
(1996).

\bibitem{Zidan2018} M. Zidan, A. Abdel, A. Younes,E. A. Zanaty, E.Khayat, M. Abdel, ``A novel algorithm based on entanglement measurement for improving speed of quantum algorithms" Appl. Math.
Inf. Sci. \textbf{12} 265 (2018).

\bibitem{Zhong2020} H. Zhong et. al, ``Quantum computational advantage using photons" Science \textbf{370}, 1460 (2020).

\bibitem{Kimble2008} H. J. Kimble, ``The quantum internet" Nature \textbf{453}, 1023, (2008).

\bibitem{Kwait2012} P. G. Kwiat, ``Hyper-entangled states" J. Mod. Opt.J. Mod. Opt. \textbf{44}, 2173 (1997).

\bibitem{Yang2005} T. Yang, Q. Zhang, J. Zhang, J. Yin, Z. Zhao,
M. Zukowski, Z. B. Chen, J. W. Pan, ``All-versus-nothing violation
of local realism by two-photon, four-dimensional entanglement"
Phys. Rev. Lett. \textbf{95}, 240406 (2005).

\bibitem{Nawaz2017} M. Nawaz, R. Islam, T. Abbas, M. Ikram, ``Engineering quantum hyperentangled states in atomic systems" J. Phys. B: At. Mol.
Opt. Phys. \textbf{50}, 215502 (2017).

\bibitem{Nawaz2018} M. Nawaz, R. Islam, T. Abbas, M. Ikram, ``Remote state preparation through hyperentangled atomic states" J. Phys. B: At. Mol. Opt.
Phys. \textbf{51}, 075501 (2018).

\bibitem{Nawaz2019} M. Nawaz, R. Islam, T. Abbas, M. Ikram, ``Atomic Cheshire cat: untying energy levels from the de Broglie motion" J. Phys. B: At. Mol.
Opt. Phys. \textbf{52}, 105501 (2018)

\bibitem{Barreiro2008} J. T. Barreiro, P. G. Kwiat, ``Hyperentanglement for advanced quantum communication" Proc. of SPIE
7092 70920P (2008).

\bibitem{Wang2016} G. Y. Wang, Q. Liu, F. G. Deng, ``Hyperentanglement purification for two-photon six-qubit quantum systems" Phys. Rev. A
\textbf{94}, 032319 (2016).

\bibitem{Wang2015} X. L. Wang, X. D. Cai, Z. F. Su, M. C. Chen, D. Wu, L.
Li, N. L. Liu, C. Y. Lu, J. W. Pan, ``Quantum teleportation of
multiple degrees of freedom of a single photon" Nature 518 516
(2015).

\bibitem{Ali2021} L. Ali, R. Islam, M. Ikram, T. Abbas and I. Ahmad, ``Hyperentanglement teleportation through external momenta states" J. P
hys. B : A t. Mol . Opt. P hys. \textbf{54} 235501(12pp) (2021).
\bibitem{Ali2022-1} L. Ali, R. Islam, M. Ikram, T. Abbas and I.
Ahmad, ``Teleportation of atomic external states on the internal
degrees of freedom" Quantum Inf. Process \textbf{21}, 55(1-15)
(2022).

\bibitem{Ali2022-2} L. Ali, R. Islam, M. Ikram, M. Imran and I.
Ahmad, ``Generation of maximally entangled N-photon field W-states
via cavity QED" Eur. Phys. J. Plus, 137, 1236(1-12) (2022).

\bibitem{Ali2024} L. Ali, M. Ahmad, R. Islam, M. Imran, M.
Ikram, I. Ahmad ``Cavity-Assisted Atomic External Momenta State
Teleportation" Ann. Phys. \textbf{536}(4), 2300277 (2024).

\bibitem{Wei2007} T. C. Wei, J. T. Barreiro and P. G. Kwiat, ``Hyperentangled Bell-state analysis" Phys. Rev.
A \textbf{75} 060305 (2007).

\bibitem{Li2015} X. H. Li, S. Ghose, ``Hyperentanglement concentration for time-bin and polarization hyperentangled photons" Phys. Rev. A \textbf{91}
062302 (2015).

\bibitem{Kunze1996} S. Kunze, S. Durr, G. Rempe, ``Bragg scattering of slow atoms from a standing light wave" Europhys. Lett.\textbf{ 34}
343 (1996).

\bibitem{Durr1998} S. Durr, T. Nonn, G. Rempe, ``Origin of quantum-mechanical complementarity probed by a which-way experiment in an atom interferometer" Nature \textbf{395} 33
(1998).

\bibitem{Islam2015} R. Islam, T. Abbas and M. Ikram, ``Biasing a coin after the toss: asymmetric delayed choice quantum eraser via Bragg regime cavity QED" Laser Phys. Lett.
\textbf{12} 015203 (2015).

\bibitem{Ikram2015} M. Ikram, M. Imran, T. Abbas and R. Islams, ``Wheeler's delayed-choice experiment: A proposal for the Bragg-regime cavity-QED implementation" Phys.
Rev. A \textbf{91} 043636 (2015).

\bibitem{Raussendorf2001} R. Raussendorf, H. J. Briegel, ``One-way quantum computation" Phys. Rev.Lett.
\textbf{86}, 5188-5191 (2001).

\bibitem{Nielsen2006} M. A. Nielsen, ``Cluster-state quantum computation" Rep. Math. Phys. \textbf{57}, 147-161
(2006).

\bibitem{Waltheri2005} P. Walther, K. J. Resch, T. Rudolph, E. Schenck, H.
Weinfurter, V. Vedral, M. Aspelmeyer, A. Zeilinger, ``Experimental
one-way quantum computing" Nature \textbf{434}, 169-176 (2005).

\bibitem{Anis2021} S. M. A. Anis, T. Abbas, M. Imran and R. Islam, ``Engineering quantum networks through Bragg diffracted hyperentangled atoms" Phys.
Scr. 96 (2021) 125102.

\bibitem{Browne2005} D. E. Browne, T.Rudolph, ``Resource-efficient linear optical quantum computation" Phys. Rev. Lett. \textbf{95},
010501 (2005).

\bibitem{Tokunaga2005} Y. Tokunaga, T. Yamamoto, M. Koashi, N. Imoto, ``Simple experimental scheme of preparing a four-photon entangled state for the teleportation-based realization of a linear optical controlled-NOT gate" Phys.
Rev. A \textbf{71}, 030301 (2005).

\bibitem{Vallone2005} G. Vallone, E. Pomarico, F. De Martini, P.
Mataloni, ``One-way quantum computation with two-photon multiqubit
cluster states" Phys. Rev. A \textbf{78}, 042335 (2008).

\bibitem{Tokunaga2008} Y. Tokunaga, S. Kuwashiro, T. Yamamoto, M. Koashi, N.
Imoto, ``Generation of high-fidelity four-photon cluster state and
quantum-domain demonstration of one-way quantum computing" Phys.
Rev. Lett. \textbf{100}, 210501 (2008).

\bibitem{Gao2010} W. B. Gao, X. C. Yao, P. Xu, H. Lu, O. Guhne, A. Cabello,
C. Y. Lu, T. Yang, Z. B. Chen, J. W. Pan, ``Bell inequality tests
of four-photon six-qubit graph states" Phys. Rev. A \textbf{82},
042334 (2010).

\bibitem{Cho2005} J. Cho, H. W. Lee, ``Generation of atomic cluster states through the cavity input-output process" Phys. Rev. Lett. \textbf{95}, 160501
(2005).

\bibitem{Dong2006} P. Dong, Z. Y. Xue, M. Yang, Z. L. Cao, ``Generation of cluster states" Phys. Rev. A
\textbf{73}, 033818 (2006).

\bibitem{Zhang2007} X. L. Zhang, K. L. Gao, M. Feng, ``Efficient and high-fidelity generation of atomic cluster states with cavity QED and linear optics" Phys. Rev. A \textbf{75}%
, 034308 (2007).

\bibitem{Lee2008} J. Lee, J. Park, S. M. Lee, H. W. Lee, A. H. Khosa, ``Scalable cavity-QED-based scheme of generating entanglement of atoms and of cavity fields" Phys.
Rev. A \textbf{77}, 032327 (2008).

\bibitem{Lee2009} G. Li, S. Ke, Z. Ficek, ``Generation of pure continuous-variable entangled cluster states of four separate atomic ensembles in a ring cavity" Phys. Rev. A \textbf{79}, 033827
(2009).

\bibitem{Gonta2009} D. Gonta, T. Radtke, S. Fritzsche, ``Generation of two-dimensional cluster states by using high-finesse bimodal cavities" Phys. Rev. A \textbf{%
79}, 062319 (2009).

\bibitem{Ballester2011} D. Ballester, J. Cho, M. S. Kim, ``Generation of graph-state streams" Phys. Rev. A
\textbf{83}, 010302(R) (2011).

\bibitem{Barrett2005} S. D. Barrett, P. Kok, ``Efficient high-fidelity quantum computation using matter qubits and linear optics" Phys. Rev. A \textbf{71},
060310(R) (2005).

\bibitem{Duan2005} L. M. Duan, R. Raussendorf, ``Efficient quantum computation with probabilistic quantum gates" Phys. Rev. Lett. \textbf{95},
080503 (2005).

\bibitem{Chen2006} Q. Chen, J. Cheng, K. L. Wang, J. Du, ``Efficient construction of two-dimensional cluster states with probabilistic quantum gates" Phys. Rev. A
\textbf{73}, 012303 (2006).

\bibitem{Tanamoto2006} T. Tanamoto, Y. X. Liu, S. Fujita, X. Hu, F.
Nori, ``Producing cluster states in charge qubits and flux qubits"
Phys. Rev. Lett. \textbf{97}, 230501 (2006).

\bibitem{Xue2006} Z. Y. Xue, G. Zhang, P. Dong,Y. M. Yi, Z. L. Cao, ``Quantum controlled phase gate and cluster states generation via two superconducting quantum interference devices in a cavity" Eur.
Phys. J. B \textbf{52}, 333-336 (2006).

\bibitem{Zhangi2006} X. L. Zhang, K. L. Gao, M. Feng, ``Preparation of cluster states and W states with superconducting quantum-interference-device qubits in cavity QED" Phys. Rev. A \textbf{74%
}, 024303 (2006).

\bibitem{Zheng2006} S. B. Zheng, ``Generation of cluster states in ion-trap systems" Phys. Rev. A \textbf{73}, 065802 (2006).

\bibitem{Blythe2006} P. J. Blythe, B. T. H. Varcoe, ``A cavity-QED scheme for cluster-state quantum computing using crossed atomic beams" New J. Phys.
\textbf{8}, 231 (2006).

\bibitem{Kiesel2005} N. Kiesel, C. Schmid, U. Weber, G. T. t~h, O. G. h~ne,
R. Ursin, H. Weinfurter, Experimental analysis of a four-qubit
photon cluster state. Phys. Rev. Lett. \textbf{95}, 210502 (2005).

\bibitem{Zhang2006}N. Zhang, C. Y. Lu, X. Q. Zhou, Y. A. Chen, Z.
Zhao, T. Yang, J. W. Pan, ``Experimental construction of optical
multiqubit cluster states from Bell states" Phys. Rev. A
\textbf{73}, 022330 (2006).

\bibitem{Hein2004}Hein, J. Eisert, H. J. Briegel, ``Multiparty entanglement in graph states" Phys. Rev. A \textbf{69}, 062311 (2004).

\bibitem{IslamR12008} Islam, A. H. Khosa, H. W. Lee, F. Saif, ``Generation of field cluster states through collective operation of cavity QED disentanglement eraser" Eur. Phys. J. D \textbf{48}, 271-277 (2008).

\bibitem{Briegel2001} H. J. Briegel, R. Raussendorf, ``Persistent entanglement
in arrays of interacting particles" Phys. Rev. Lett. \textbf{86},
910 13 (2001).

\bibitem{Briegel2004} D. R. W. Briegel, ``Stability of macroscopic entanglement under decoherence" J. Phys. Rev. Lett. \textbf{92},
180403 (2004).

\bibitem{Zou2007} X. Zou, Y. Xiao, S. Li, Y. Yang and G. Guo,
``Quantum phase gate through a dispersive atom-field interaction"
Phys. Rev. A, \textbf{75}, 064301 (2007).

\bibitem{Khosa2004} A. H. Khosa, M. Ikram, M. S. Zubairy, ``Measurement of entangled states via atomic beam deflection in Bragg's regime" Phys. Rev.
A \textbf{70}, 052312 (2004).

\bibitem{Khosa2005} A. H. Khosa, M. S. Zubairy, ``Quantum-state measurement of two-mode entangled field-state in a high- Q cavity" Phys. Rev. A \textbf{72}%
, 42106 (2005).

\bibitem{Khosa2006} A. H. Khosa, M. S. Zubairy, ``Measurement of Wigner function via atomic beam deflection in the Raman-Nath regime" J. Phys. B At.
Mol. Opt. Phys. \textbf{39}, 5079 --5~089 (2006).

\bibitem{Khan1999} A. A. Khan, M. S. Zubairy, ``Quantum non-demolitionmeasurement of Fock states via atomic scattering in Bragg regime" Phys. Lett. A \textbf{254}, 301 --3~06 (1999).

\bibitem{Khalique2003} A. Khalique, F. Saif, ``Engineering entanglement between external degrees of freedom of atoms via Bragg scattering" Phys.
Lett. A, \textbf{314}, 37-43 (2003).

\bibitem{Islam2007} R. Islam, M. Ikram, F. Saif, ``Engineering maximally entangled N-photon NOON field states using atom interferometer based on Bragg regime cavity QED" J. Phys. B \textbf{40}, 1359 1-368 (2007).

\bibitem{Islam2008} R. Islam, A. H. Khosa, F. Saif, ``NOON and W states via atom interferometry" J. Phys. B \textbf{41}, 035505
(2008).

\bibitem{Qamar2003} S. Qamar, S. Y. Zhu, M. S. Zubairy, ``Teleportation of an atomic momentum state" Phys. Rev. A \textbf{67}, 042318 (2003).

\bibitem{Islam2009} R. Islam, M. Ikram, R. Ahmed, A. H. Khosa, F. Saif, ``Atomic state teleportation, from internal to external degrees of freedom" J. Mod. Opt. \textbf{56}, 875 ~80 (2009).

\bibitem{Scully1997} M. O. Scully and M. S. Zubairy, Quantum Optics" Cambridge University
Press, Cambridge 1997.
\bibitem{Haroche2006}  S. Haroche, J. M. Raimond, ``Exploring the Quantum: Atoms, Cavities and Photons" Oxford University Press,
Oxford 2006.
\bibitem{Haroche2020}   S. Haroche, J.M. Raimond, ``From cavity to circuit quantum electrodynamics" Nat. Phys. \textbf{16} 243-246
(2020).
\bibitem{Zubairy2004} M. S. Zubairy, G. S. Agarwal, M. O. Scully, ``Quantum disentanglement eraser: a cavity QED implementation" Phys.
Rev. A \textbf{70} 012316 (2004).
\bibitem{Rauschenbeutel1999} A. Rauschenbeutel, G. Nogues, S. Osnaghi, P.
Bertet, M. Brune, J. M. Raimond, S. Haroche, ``Coherent operation
of a tunable quantum phase gate in cavity QED" Phys. Rev. Lett.
\textbf{83}, 5166 --5~169 (1999).

\bibitem{Cohen2018} I. Cohen and K. Molmer ``Deterministic
Quantum Network for Distributed Entanglement and Quantum
Computation" Phys. Rev. A, \textbf{98}, 030302 (2018).

\bibitem{Ball2008} P. Ball, ``Physics: Quantum all the way" Nature \textbf{453} 22 (2008).

\bibitem{RAH1}A. U. Rahman, M. Noman, M. Javed, and A. Ullah, Eur. Phys. J. Plus, \textbf{136(8)},
1-19 (2021).

\bibitem{RAH2}A. U. Rahman, M. Noman, M. Javed, M. X. Luo, and A. Ullah, Quantum Inf. Process. \textbf{20(9)},
1-20 (2021).
\bibitem{RAH3} A. U. Rahman, M. Javed, and A. Ullah, Probing multipartite entanglement, coherence and quantum
information preservation under classical Ornstein-Uhlenbeck noise.
arXiv preprint arXiv:2107.11251.

\bibitem{RAH4} A. U. Rahman, M. Javed, A. Ullah, and Z. X. Ji, Quantum Inf. Process. \textbf{20}, 321 (2021).
https://doi.org/10.1007/s11128-021-03257-z.

\bibitem{Ali2023} L. Ali, A. U. Rahman, M. Imran, R. Islam, M.
Ikram and I. Ahmad ``The infuence of mixed classical dephasing
noisy channels on the dynamics of two-qubit correlations" Opt.
Quantum Electron.\textbf{ 55}, 120 (2023).

\bibitem{Elliott2002}C. Elliott ``Building the quantum network"
New J. Phys. \textbf{4}, 46 (2002).

\bibitem{Simon2017} C. Simon ``Towards a global quantum network" Nat. Photonics, \textbf{11}, 678-680 (2017).

\bibitem{Lloyd2011} S. Lloyd ``Quantum coherence in biological
systems" J. Phys. Conf. Ser. \textbf{302}, 012037 (2011).

\bibitem{Lambert2013} N. Lambert, Y. Chen, Y. Cheng, C. Li, G.
Chen and F. Nori ``Quantum biology" Nat. Phys. \textbf{9}, 10-18
(2013).

\bibitem{Gadiyaram2019} V. Gadiyaram, S. Vishveshwara and S.
Vishveshwara ``From Quantum Chemistry to Networks in Biology: A
Graph Spectral Approach to Protein Structure Analyses" J. Chem.
Inf. Model. \textbf{59(5)}, 1715-1727 (2019).

\bibitem{Kim2021} Y. Kim, F. Bertagna, E. M. Souza, D. J. Heyes,
L. O. Johannissen, E. T. Nery, A. Pantelias, A. S. Jimenez, L.
Slocombe, M. G. Spencer, J. Al-Khalili, G. S. Engel, S. Hay, S. M.
Wilson, K. Jeevaratnam, A. R. Jones, D. R. Kattnig, R. Lewis, M.
Sacchi, N. S. Scrutton, S. R. P. Silva and J. McFadden ``Quantum
Biology: An Update and Perspective" Quant. Reports, \textbf{3},
1-48 (2021).

\bibitem{McCutcheon2016}  W. McCutcheon, A. Pappa, B. A. Bell, A. McMillan, A. Chailloux, T. Lawson, M. Mafu, D. Markham, E. Diamanti, I. Kerenidis, J. G. Rarity and M. S. Tame, ``Experimental verification of multipartite entanglement in quantum networks" Nat. Comm. \textbf{7} 13251 (2016).
\bibitem{Cavalcanti2015}   D. Cavalcanti, P. Skrzypczyk, G. H. Aguilar, R. V.
Nery, P. H. Souto Ribeiro, S. P. Walborn, ``Detection of
entanglement in asymmetric quantum networks and multipartite
quantum steering" Nat. Comm. \textbf{6} 7941 (2015).

\bibitem{Liao2018} S. Liao et.al, ``Satellite-relayed
intercontinental quantum network" Phys. Rev. Lett. \textbf{120},
030501 (2018).

\bibitem{Dynes2019}J. F. Dynes, A. Wonfor, W. W. S. Tam, A. W.
Sharpe, R. Takahashi, M. Lucamarini, A. Plews, Z. L. Yuan, A. R.
Dixon, J. Cho, Y. Tanizawa, J. P. Elbers, H. Greiber, I. H. White,
R. V. Penty and A. J. Shields ``Cambridge quantum network" Nat.
Photonics, J. Quant. Inf. \textbf{5}, 101 (2019).

\bibitem{Ball2011} P. Ball ``Physics of life: The dawn of
quantum biology" Nature, \textbf{474}, 272-274 (2011).

\bibitem{Arndt2009}M. Arndt, T. Juffmann and V. Vedral ``
Quantum physics meets biology" HFSP J. \textbf{3}, 386-400 (2009).

\bibitem{Cirac1999}J. I. Cirac, A. K. Ekert, S. F. Huelga and C.
Macchiavello ``Distributed Quantum Computation over Noisy
Channels" Phys. Rev. A, \textbf{59}, 4249 (1999).

\bibitem{Cortinas2020}  R. G. Cortinas,  M. Favier, B. Ravon, P. Mehaignerie,
  Y. Machu, J. M. Raimond, C. Sayrin, M. Brune, ``Laser trapping of circular Rydberg atoms" Phys. Rev. Lett. \textbf{124}
123201 (2020).
\bibitem{Meystre2021}  P. Meystre, Quantum optics: taming the
quantum. Springer, Switzerland (2021).

\bibitem{Deleglise2008} S. Deleglise, I. Dotsenko, C. Sayrin
J. Bernu,  M. Brune, J. M. Raimond,  S. Haroche, ``Reconstruction
of non-classical cavity field states with snapshots of their
decoherence" Nature \textbf{455} 510 (2008).
\bibitem{Gleyzes2007} S. Gleyzes, S. Kuhr,  C. Guerlin, J.
Bernu, S. Deleglise,  U. B. Hoff, S. Haroche, ``Observing the
quantum jumps of light: birth and death of a photon in a cavity"
Nature \textbf{446} 297 (2007).

\bibitem{Giltner1995} D. M. Giltner,  R. W. McGowan, S. A. Lee, ``Theoretical and experimental study of the Bragg scattering of atoms from a standing light wave" Phys. Rev.
A \textbf{52} 3966 (1995).
\bibitem{Martin1087}   P. J. Martin,   P. L. Gould,  B. G.Oldaker,  A. H. Miklich
D. E. Pritchard, ``Diffraction of atoms moving through a standing
light wave" Phys. Rev. A \textbf{36} 2495 (1987).
\bibitem{Vernooy1998}  D. W. Vernooy,
 V. S. Ilchenko,  H. Mabuchi,  E. W. Streed H. J. Kimble, ``High-Q measurements of fused-silica microspheres in the near infrared" Opt. Lett.
\textbf{23} 247 (1998).

\bibitem{Durr1996}  S. Durr,  S. Kunze,  G. Rempe, ``Pendellosung oscillations in second-order Bragg scattering of atoms from a standing light wave" Quantum Semiclassical Opt. \textbf{8}
531-539 (1996).
\bibitem{Munstermann1999}   P. Munstermann, T. Fischer, P. W. H. Pinkse, G. Rempe, ``Single slow atoms from an atomic fountain observed in a
high-finesse optical cavity" Opt. Commun. \textbf{159} 63-67
(1999).
\bibitem{Munstermanni1999}   P. Munstermann, T. Fischer,  P. Maunz, P. W. H. Pinkse,
G. Rempe, ``Dynamics of single-atom motion observed in a
high-finesse cavity" Phys. Rev. Lett. \textbf{82} 37913797 (1999).
\bibitem{Puppe2004}  T. Puppe,  P. Maunz,  T. Fischer,  P. W. H. Pinkse,
 G. Rempe, ``Single-atom trajectories in higher-order transverse modes of a high-finesse optical cavity" Phys. Scr. T \textbf{112} 7
(2004).
\bibitem{Hood2001} C. J. Hood,  H. J. Kimble,  J. Ye, ``Characterization of high-finesse
mirrors: Loss, phase shifts, and mode structure in an optical
cavity" Phys. Rev. A \textbf{64} 033804 (2001).
\bibitem{Mohseni2014}   M. Mohseni, Y. Omar, G. S. Engel, M. B. Plenio, ``Quantum Effects in Biology" Cambridge University Press, Cambridge 2014.
\bibitem{Aharonov2018}  Y. Aharonov, E. Cohen, J. Tollaksen, ``Completely top-down hierarchical structure in quantum mechanics" PNAS, \textbf{115}
11730 (2018).

\end{thebibliography}
\end{document}